\documentclass[a4paper,11pt]{article}
\pdfoutput=1
\usepackage{jheppub} 
\usepackage[utf8]{inputenc}
\usepackage{amssymb}
\usepackage{amsmath}
\usepackage{amsfonts}
\usepackage{graphicx}
\usepackage{color}
\usepackage{xspace}
\usepackage{ulem}
\usepackage{lscape}
\usepackage{ wasysym }

\usepackage{float}
\usepackage{subfig}
\usepackage{mathrsfs}
\usepackage{slashed}

\usepackage{multirow,rotating}

\def\gsim{\raise0.3ex\hbox{$\;>$\kern-0.75em\raise-1.1ex\hbox{$\sim\;$}}}
\def\lsim{\raise0.3ex\hbox{$\;<$\kern-0.75em\raise-1.1ex\hbox{$\sim\;$}}}

\newcommand{\ba}[1]{\begin{eqnarray} \label{(#1)}}
\newcommand{\ea}{\end{eqnarray}}

\newcommand{\AddrAHEP}{
  {\it AHEP Group, Instituto de F\'{\i}sica Corpuscular --
    C.S.I.C./Universitat de Val{\`e}ncia \\
    Edificio de Institutos de Paterna, Apartado 22085,
  E--46071 Val{\`e}ncia, Spain}}
  
  \newcommand{\AddrUFSM}{
Departamento de F\' isica, Facultad de Ciencias, Universidad de La Serena, \\
Avenida Cisternas 1200, La Serena, Chile.  \\
Centro-Cient\'\i fico-Tecnol\'{o}gico de Valpara\'\i so, \\ 
Casilla 110-V, Valpara\'\i so,  Chile.
}

\newcommand{\AddrBonn}{%
Bethe Center for Theoretical Physics \& Physikalisches Institut der 
Universit\"at Bonn,\\ Nu{\ss}allee 12, 
 53115 Bonn, Germany
}

\def\gsim{\raise0.3ex\hbox{$\;>$\kern-0.75em\raise-1.1ex\hbox{$\sim\;$}}}
\def\lsim{\raise0.3ex\hbox{$\;<$\kern-0.75em\raise-1.1ex\hbox{$\sim\;$}}}

\begin{document}


\title{Heavy neutral fermions at the high-luminosity LHC}

\author[a]{Juan Carlos Helo,} \emailAdd{jchelo@userena.cl}\affiliation[a]{\AddrUFSM}
\author[b]{Martin Hirsch} \emailAdd{mahirsch@ific.uv.es}\affiliation[b]{\AddrAHEP}
\author[c]{and Zeren Simon Wang}\emailAdd{wzeren@physik.uni-bonn.de}\affiliation[c]{\AddrBonn}
\abstract{
  Long-lived light particles (LLLPs) appear in many extensions of the
  standard model. LLLPs are usually motivated by the observed small
  neutrino masses, by dark matter or both. Typical examples for
  fermionic LLLPs (a.k.a. heavy neutral fermions, HNFs) are sterile
  neutrinos or the lightest neutralino in R-parity violating
  supersymmetry. The high luminosity LHC is expected to deliver up to
  3/ab of data. Searches for LLLPs in dedicated experiments at the LHC
  could then probe the parameter space of LLLP models with
  unprecedented sensitivity.  Here, we compare the prospects of
  several recent experimental proposals, FASER, CODEX-b and MATHUSLA,
  to search for HNFs and discuss their relative merits.
  }


\arxivnumber{1803.02212}

\vskip10mm

\maketitle
\flushbottom
%

\section{Introduction}
\label{sect:intro}

Light long-lived particles (LLLPs) appear in many extensions of the
standard model (SM). LLLPs can be scalars, fermions or vectors. Fermionic
LLLPs are also often called heavy neutral fermions (HNF) in the
literature. LLLPs are usually motivated either by dark matter, by
small neutrino masses or by both. A discussion of different
theoretical models for LLLPs can be found, for example in
\cite{Essig:2013lka,Alekhin:2015byh}.

In the past few years several experimental proposals with improved
sensitivities to LLLPs have been discussed. For example, there are the
planned fixed target experiment SHiP \cite{Alekhin:2015byh}, the near
detector of the future DUNE experiment \cite{Adams:2013qkq}, or also
NA62 \cite{NA62:2017rwk}. However, note that the primary goal of NA62
is to measure precisely Br($K^+\to \pi^+ \nu {\bar \nu}$), while the
main task of the near detector of DUNE is just monitoring the neutrino
flux for the far detector \cite{Adams:2013qkq}.

It is expected that the LHC will deliver up to ${\cal L}=3000/$fb of
luminosity over the next (15-20) years \cite{High-Luminosity:2114693}.
Perhaps unsurprisingly, a number of new proposals to search for LLLPs
have appeared, all based on the idea to exploit LHC's large
luminosity: MATHUSLA \cite{Chou:2016lxi}, CODEX-b
\cite{Gligorov:2017nwh} and FASER \cite{Feng:2017uoz}. The physics
potential of these three experiments has so far not been fully
discussed in the literature and it is the aim of the current paper to
estimate, and compare to each other and previous experiments, the
sensitivity of these proposals for fermionic LLLPs.

MATHUSLA \cite{Chou:2016lxi} is a proposed very massive detector,
possibly to be located above ground on top of the ATLAS experiment.
The sizeable distance of MATHUSLA from the interaction point (IP) of
the LHC beams will allow to test for rather long life-times, up to
$c\tau \lsim 10^{(7-8)}$ m, but requires a huge detection volume.
CODEX-b \cite{Gligorov:2017nwh} is a proposal that takes advantage of
a relatively large shielded empty space near the location of the LHCb
experiment. Being closer to the IP, CODEX-b proposed size is much
smaller than that of MATHUSLA. Finally, the authors of FASER
\cite{Feng:2017uoz} propose to construct a very modestly sized
detector, situated in the very forward direction, close to either the
ATLAS or the CMS IP. Note that FASER is discussed in three different
variants in \cite{Feng:2017uoz,Feng:2017vli}. We will summarize the
experimental parameters of the different FASER setups and those of
CODEX-b and MATHUSLA in section \ref{sect:exp}.

In the original paper on the MATHUSLA detector \cite{Chou:2016lxi},
exotic Higgs decays to LLLPs were used to study the sensitivity of the
experiment. Similarly, the authors of CODEX-b \cite{Gligorov:2017nwh}
used a Higgs portal model for LLLPs to investigate the reach of their
proposal.  They also considered \cite{Gligorov:2017nwh} a light
neutral scalar, that mixes with the Higgs, produced in B-mesons decays
as a LLLP candidate.  The original FASER \cite{Feng:2017uoz}
publication studied dark photons produced in meson decays to estimate
the sensitivity of the different FASER setups.  Here, we study the
reach of these three experimental proposals for the case that the LLLP
is a heavy neutral fermion (HNF). We concentrate on two particular
example models of HNFs: (i) Sterile neutrinos and (ii) the lightest
neutralino in R-parity violating supersymmetry. As discussed in
section \ref{setc:models} both models are motivated by being possible
explanations for the observed small neutrino masses (and mixings).  In
our simulation we consider all possible LLLP production channels:
D-mesons, B-mesons, W and Z bosons, as well as Higgs boson. We
compare the different experiments systematically for the different
channels and then discuss sensitivities for our example models.

We note that, very recently in \cite{Kling:2018wct} the sensitivity of
FASER to sterile neutrinos was estimated. We will comment on this work
in more detail later, but note here only briefly that our estimates
roughly, but not completely, agree with those given in
\cite{Kling:2018wct}.  However, \cite{Kling:2018wct} studies only the
case of sterile neutrinos, while we also discuss R-parity violating
neutralinos, and concentrates exclusively on FASER, while we take into
account the different experimental proposals discussed above and also
compare to the beam-dump experiments
\cite{Adams:2013qkq,Alekhin:2015byh,NA62:2017rwk}.

Before closing this introduction, we mention that there exist of
course already many searches for sterile neutrinos (and other HNFs).
For a review on constraints for sterile neutrino see, for example
\cite{Atre:2009rg}.  Also the main LHC experiments, ATLAS and CMS,
have searched for HNFs. ATLAS \cite{Aad:2015xaa} published results of
a search based on the final state $lljj$, giving only weak upper
limits on the mixing of the sterile neutrinos $V_{\alpha N}^2\simeq
(10^{-2}-10^{-1})$ ($\alpha=e,\mu$) for $m_N \sim (100-500)$ GeV.  CMS
searched for sterile neutrinos in trilepton final states and very
recently published limits as low as $V_{\alpha N}^2 \simeq 10^{-5}$ in
the mass range ($10-100$) GeV \cite{Sirunyan:2018mtv}. With these
results \cite{Sirunyan:2018mtv}, CMS now gives limits competitive with
those derived by the DELPHI experiment at LEP \cite{Abreu:1996pa}.

Note that for small mixing angles $V_{\alpha N}^2$ below, say
$V_{\alpha N}^2 \sim 10^{-7}$, for $m_N \sim {\cal O}(10)$ GeV, the
decay lengths of sterile neutrinos become large enough to be detected
exerpimentally and ATLAS/CMS could search for sterile neutrinos using
the ``displaced vertex'' signal \cite{Helo:2013esa}.  However, current
displaced vertex search strategies, as used by CMS
\cite{Sirunyan:2017jdo} for example, are not very well suited for
light, say $m_N \lsim 100$ GeV, sterile neutrinos
\cite{Cottin:2018kmq}.

We also mention in passing that our other HNF candidate, the
neutralino, has been studied as an LLLP candidate before.  R-parity
violating SUSY and neutralinos as LLLPs are mentioned in the SHiP
proposal \cite{Alekhin:2015byh} and the SHiP sensitivity for
neutralinos has been studied in more details in
\cite{Gorbunov:2015mba,deVries:2015mfw}.

The rest of this paper is organized as follows. In section
\ref{setc:models} we briefly summarize the basics of sterile neutrinos
and neutralinos with R-parity violation. In section \ref{sect:exp} we
give a basic description of the different experiments and outline our
simulations. In section \ref{sect:res} we discuss our numerical
results, before closing with a short summary.

\section{Heavy neutral fermions: Example models}
\label{setc:models}

In this section we give a short summary of the two models for heavy
neutral fermions that we study numerically in this paper.  We first
discuss sterile neutrinos and then give some basic definitions and
features for the neutralino in R-parity violating supersymmetry. Since
both models have been discussed in the literature many times, we will
be very brief.

\subsection{Sterile neutrinos}
\label{subsect:sterile}

The standard model predicts neutrinos to be massless, in contrast
to the results of neutrino oscillation experiments.
\footnote{For the status of oscillation data, see for example the
  recent global fit \cite{deSalas:2017kay}.}  The simplest extension
of the SM, which can explain the experimental data, adds $n$ fermionic
singlets. Oscillation data requires $n \ge 2$. The Lagrangian of this
model contains two new terms
\begin{equation}\label{eq:lag}
{\cal L}^{\nu_R} = Y_{\nu} {\bar L} H^{\dagger} \nu_R + M_N \nu_R \nu_R
\end{equation}
Here, we have suppressed generation indices. In general $M_N$ is a
complex symmetric ($n,n$) matrix, while $Y_{\nu}$ is a ($3,n$)
matrix. In the simple model considered here, without new interactions
for the $\nu_R$, one can perform a basis change and choose the entries
of $M_N$ to be diagonal, real and positive. The masses of the active
neutrinos are small, if $(Y_{\nu}v)\cdot M_N^{-1} \ll 1$, this is the
essence of the seesaw mechanism. Diagonalization of the mass matrix
leads then to three light, active neutrinos and $n$ nearly sterile
mass eigenstates, which we denote by $\nu_S$ in the following.

The heavy sterile neutrino Charged (CC) and Neutral
Current (NC) interactions are
\begin{eqnarray}\label{CC-NC}
{\cal L} &=& \frac{g}{\sqrt{2}}\, 
 V_{\alpha N_j}\ \bar l_\alpha \gamma^{\mu} P_L \nu_{S_j} W^-_{L \mu} +\frac{g}{2 \cos\theta_W}\ \sum_{\alpha, i, j}V^{L}_{\alpha i} V_{\alpha N_j}^*  
\overline{\nu_{S_j}} \gamma^{\mu} P_L \nu_{i} Z_{\mu},
\end{eqnarray}
where $i=1,2,3$ and $j=1, .., n$ and $\alpha$ denotes the charged lepton
generation. The left-handed sector neutrino mixing matrix $V^{L}$ is
measured in neutrino oscillations. $V_{\alpha N_j}$ describes the
mixing between ordinary and sterile neutrinos. Within the simple
seesaw model, described by eq. (\ref{eq:lag}), one expects that
$V_{\alpha N_j}$ is roughly of the order of $V_{\alpha N_j} \propto
\sqrt{m_{\nu}/M_N}$, i.e. $|V_{\alpha N_j}|^2 \simeq 5 \times 10^{-11}
(\frac{m_{\nu}}{\rm 0.05 eV})(\frac{\rm 1 GeV}{M_N})$.  However, in
extensions of this simple framework, for example the inverse
seesaw \cite{Mohapatra:1986bd}, much larger values for the mixing can
occurr, despite the smallness of the observed neutrino masses.  For
this reason, for the sensitivity estimates of the different
experiments we will take $|V_{\alpha N_j}|^2$ as a free parameter in
our calculations.  Note that the mixing between sterile and active
neutrinos controls both, production and decay of the sterile states.

Oscillation data shows two large mixing angles in the active neutrino
sector \cite{deSalas:2017kay}. Thus, one expects that the heavy
sterile neutrinos couple typically to more than one generation of
charged leptons too, see eq. (\ref{CC-NC}). It is easy to fit all
oscillation data with the seesaw mechanism, described by
eq. (\ref{eq:lag}).  However, the Yukawa matrices can be fixed by such
a fit only up to an orthogonal rotation matrix containing three
complex parameters \cite{Casas:2001sr}, leaving $|V_{\alpha N_j}|$
essentially as free parameters.\footnote{For an extension of this
  Casas-Ibarra parametrization for the inverse seesaw case, see
  \cite{Anamiati:2016uxp}.} In our sensitivity estimates we will
simply assume that only one sterile neutrino exists in the mass range
to which the experiments are sensitive. We will also not distinguish
between $e$ and $\mu$ flavours, assuming simply that only one of the
corresponding $|V_{\alpha N}|$ is non-zero. Since we are only
interested in estimating sensitivity ranges, not in a full
reconstruction of the seesaw parameters, this should be a reasonable
approximation.

\subsection{R-parity violating neutralino}
\label{subsect:rpv}

Supersymmetric models with R-parity violation (RPV) have been
discussed in detail in many publications in the literature, for
reviews see for example \cite{Barbier:2004ez,Dreiner:1997uz}. R-parity
violating terms do either break baryon ($B$) or lepton ($L$) number.
The lepton number violating (LNV) part of the superpotental can 
be written as:
\begin{equation}\label{eq:WpotL}
W = \lambda_{ijk} {\hat L}_i {\hat L}_j {\hat E}^c_K 
+ \lambda'_{ijk} {\hat L}_i {\hat Q}_j {\hat D}^c_K 
+ \epsilon_i  {\hat L}_i {\hat H}_u,
\end{equation}
while the baryon number violating (BNV) terms are:
\begin{equation}\label{eq:WpotB}
W =  \lambda''_{ijk} {\hat U^c}_i {\hat D^c}_j {\hat D}^c_K .
\end{equation}
It is experimentally excluded that both terms are present at the same
time, since otherwise the proton decays with an unacceptable rate
unless the product of the two couplings is tiny \cite{Dreiner:1997uz},
typically $\lambda' \times \lambda'' \lsim 10^{-24}$ for TeV SUSY
masses.  The lepton number violating terms generate neutrino masses
and it is well-known that even the simplest bilinear RPV model can fit
oscillation data \cite{Hirsch:2000ef}. We will therefore put all BNV
terms to zero in the following. Once R-parity is violated, the
lightest SUSY particle is no longer stable and therefore there are no
constraints on its nature from cosmology. Thus, even charged or
coloured SUSY particles could be the LSP (lightest supersymmetric
particle). Here, we are exclusively interested in the case where the
lightest neutralino is the LSP.

In the so-called CMSSM the gaugino mass terms $M_1$ and $M_2$ follow
the approximate relation $M_1 \simeq (1/2) M_2$. This leads to a lower
limit on $m_{\chi^0_1}$ of roughly $m_{\chi^0_1} \gsim 46$ 
GeV \cite{Patrignani:2016xqp}. However, in more general SUSY 
models, $M_1$ and $M_2$ are just free parameters and it is easy 
to show that for \cite{Dreiner:2009ic}
\begin{equation}\label{eq:lightNtrl}
M_1 = \frac{M_2M_Z^2\sin(2\beta)\sin^2\theta_W}
{\mu M_2 -M_Z^2\sin(2\beta)\cos^2\theta_W}
\end{equation}
the lightest neutralino is massless at tree-level. In our numerical
studies we will simply take the mass of the lightest neutralino,
$m_{\chi^0_1}$ as a free parameter, without resorting to any
underlying SUSY breaking model. Note, however, that this lightest
neutralino necessarily has to be mostly bino, due to the lower
mass limits on charginos that can be derived from LEP data
\cite{Patrignani:2016xqp}.

Decays of the lightest neutralino can be induced via either bilinear
or trilinear RPV terms.  In the former case, the neutralinos and the
neutrinos mix at tree-level, thus the neutralino can decay via
diagrams involving $W$ and $Z$ bosons.  A rough guess for the width of
the neutralino in BRPV (bilinear RPV) can be derived from the results of
\cite{Porod:2000hv}.  An order of magnitude estimate can be
given as:
\begin{equation}\label{eq:GamNtrlBRPV}
  \Gamma(\chi^0_1) \sim \frac{g^4}{512 \pi^3}
      \Big(\frac{m_{\nu}}{m_{SUSY}}\Big)\Big(\frac{m_{\chi^0_1}^5}{m_W^4}\Big)
\end{equation}
Typical decay lengths of order ${\cal O}(10-100)$ m result for
$m_{\chi^0_1} \simeq 40$ GeV for a value of $m_{\nu} \simeq
\sqrt{\Delta m_{\rm Atm}^2}$, where $\Delta m_{\rm Atm}^2$ is the
atmospheric neutrino oscillation mass scale, and SUSY masses in the
range ($100-1000$) GeV, while for $m_{\chi^0_1} \simeq 5$ GeV one
expects decay lengths of the order of ${\cal O}(100-1000)$ km. These
decay lengths fit nicely into the range of sensitivities of CODEX-b,
FASER and particular MATHUSLA, see the discussion in section
\ref{subsect:ntrl}.

Contributions to the decay of the neutralino from trilinear RPV terms,
on the other hand, involve sfermion exchange diagrams.  The order of
magnitude of the decay width of the neutralino can be estimated in
this case as:
\begin{equation}\label{eq:GamNtrl}
  \Gamma(\chi^0_1 \to l jj) \sim
  (\lambda')^2\frac{\Big( g\tan\theta_W\Big)^2}{512 \pi^3}
  \frac{m_{\chi^0_1}^5}{m_{\tilde f}^4}.
\end{equation}
Here, $\lambda'$ stands symbolically for any of the trilinear
couplings in eq. (\ref{eq:WpotL}) and $m_{\tilde f}$ is the
corresponding scalar fermion mass (either squark or slepton). Given
that there are currently only lower limits on all sfermion masses, it
is possible to take either trilinear couplings and the sfermion masses
or simply the total width of the $\chi^0_1$ as a free parameter in
trilinear RPV.

At the LHC, neutralinos can either be pair produced, via diagrams
involving either a Z-boson or a Higgs, or singly produced via R-parity
violating couplings. Since one expects R-parity violating couplings to
be small parameters
\cite{Barbier:2004ez,Dreiner:1997uz,Hirsch:2000ef}, we will focus on
pair production.  $Z$ bosons are produced much more abundantly in the
LHC than the Higgs. We will therefore concentrate the following
discussion on Z-bosons.  $Z$ bosons can decay to pairs of neutralinos,
if $m_{\chi_1^0}\lsim m_Z/2$.  The decay width $\Gamma(Z\rightarrow
\chi_1^0 \chi_1^0)$ has been calculated in
\cite{Bartl:1988cn}. Important for us is the coupling between
$Z$-boson and two neutralinos:
\begin{equation}\label{eq:Znn}
  g_{Z\chi^0_i\chi^0_j} = (N_{i3}N_{j3}-N_{i4}N_{j4}) c_{2\beta}
  + (N_{i3}N_{j4}+N_{i4}N_{j3}) s_{2\beta} .
\end{equation}  
Here, $N_{ij}$ is the matrix that diagonalizes the neutralino mass
matrix and $c_{2\beta}/s_{2\beta}$ are cosine and sine of $\beta$,
with $\tan\beta$ being the usual ratio of the vacuum expectation
values of the two Higgs doublets. Thus, the relevant coupling for the
decay to neutralinos is proportional to the Higgsino content in
$\chi^0_1$ and is {\em not suppressed by small RPV parameters}.
Different from the case of the sterile neutrino, therefore for
neutralinos production cross section and decay length {\em are not
  related}.  This important distinction between our two LLLP
candidates will be discussed in more detail in section
\ref{subsect:ntrl}.

Importantly, there is an upper bound on $\Gamma(Z\rightarrow \chi_1^0
\chi_1^0)$ from the LEP measurement of the invisible width of the
Z-boson.  The PDG \cite{Patrignani:2016xqp} gives for
$\Gamma(Z\rightarrow$inv$)$ a value in agreement with the standard
model calculation with three generations of light neutrinos and the
error bar on the measurement corresponds to an upper limit on the
branching ratio into additional invisibly final states of roughly
Br($Z\rightarrow\chi^0_i\chi^0_j$) $\lsim 0.1$ \% at 90 \% c.l.

The Higgsino content in the lightest neutralino depends mostly on the
parameter combination $M_1/\mu$. LEP gives a lower limit of roughly
$\mu \gsim 100$ GeV\cite{Patrignani:2016xqp}. A recent LHC search in
ATLAS for electroweak SUSY production \cite{Aaboud:2017leg} excludes
now values of $\mu$, depending on other SUSY parameters, up to roughly
$\mu \simeq 130$ GeV. For a cross check, we used the model
\textit{MSSMTriRpV} from the repository of \texttt{SARAH}-4.12.2
\cite{Staub:2013tta}. We perform numerical calculations with
\texttt{SPheno}-4.0.3 \cite{Porod:2011nf,Porod:2003um}.  For the choice
\begin{eqnarray}
M_2=500 \text{~GeV}, \mu =130 \text{~GeV}, 
\tan{\beta}=10, 
\end{eqnarray}
and a lightest neutralino mass $m_{\chi^0_1}\ll m_Z/2$, we find
Br($Z\rightarrow\chi^0_i\chi^0_j$) $\simeq 0.06$ \%. Thus, given
current constraints on SUSY parameters, the Higgsino content in the
lightest neutralino can still be large enough to (nearly) saturate the
experimental bound on Br($Z\rightarrow\chi^0_i\chi^0_j$). 
\footnote{We have also checked that such a ``largish'' Higgsino
  content is not in disagreement with the experimental upper
  bound on the Higgs invisible width
  \cite{Aad:2015txa,Aaboud:2017bja}.}  

In our numerical calculations, see section \ref{subsect:ntrl}, we will
not do a scan over the soft SUSY breaking parameters. Instead we will
treat both, the mass of the lightest neutralino and
Br($Z\rightarrow\chi^0_1\chi^0_1$) as a free parameter in our
numerical study. Note, however, that a future lower limit on $\mu$
larger than the numbers quoted above will consequently result in
smaller values for the maximally achievable
Br($Z\rightarrow\chi^0_1\chi^0_1$). We would also like to mention that
the lower limits on charged SUSY particles require that the lightest
neutralino must be mostly bino, for the low mass we consider in
section  \ref{subsect:ntrl}.

\section{Experimental setups \& simulation} 
\label{sect:exp}

Here we first give a brief description of the different experiments
considered. For more details we refer to the original publications for
MATHUSLA \cite{Chou:2016lxi}, CODEX-b \cite{Gligorov:2017nwh} and
FASER \cite{Feng:2017uoz}.  We then describe our numerical simulation
of these proposals. Note that, given that the experimental setups
might still undergo some changes and refinements we have not aimed at
very high accuracy in our simulations of the experiments, although we
have checked -- wherever possible -- that our calculations agree with
results obtained in previous works, see also section \ref{sect:res}.

CODEX-b (``Compact detector for Exotics at LHCb'')
\cite{Gligorov:2017nwh} was proposed recently as a detector for LLLPs.
The experiment consists of a cubic box with approximate dimensions of
($10 \times 10 \times 10$) m, installed in a free space near the LHCb
experiment. As discussed in \cite{Gligorov:2017nwh}, with some modest
amount of additional shielding CODEX-b is expected to operate in the
low background environment, necessary to search for very rare events.

The FASER (``ForwArd Search ExpeRiment'') proposal uses a cylindrical
detector a few hundred meters downstream of the ATLAS or CMS IP. The
FASER papers \cite{Feng:2017uoz,Feng:2017vli,Kling:2018wct} discuss
several different options for the position and size of the
detector. We use the following three options given in
\cite{Feng:2017vli}: FASER$^{r}$, the small FASER with radius $0.2$ m,
FASER$^{R}$ (large Radius) with radius of $1$ m and FASER$^n$
(``near'') with small radius $0.04$ m at a shorter distance.  Note
that especially for FASER$^n$, backgrounds from the LHC beam might be
a concern, since there is rather little shielding in the FASER near
position.

MATHUSLA (``MAssive Timing Hodoscope for Ultra Stable neutraL
pArticles'') \cite{Chou:2016lxi} proposes to put a surface detector
above the ATLAS interaction point. With a distance between $140-320$ m
from the IP, the MATHUSLA detector has to be quite massive, compared
to the other two proposals. Its dimensions are given as
\cite{Chou:2016lxi} $200m\times 200m\times 20m$. Being above ground,
cosmic rays are a serious background concern.  The authors of
\cite{Chou:2016lxi} discuss in some detail, how this background can be
kept under control, surrounding the decay volume with scintillator
detectors and Resistive Plate Chambers (RPCs). The excellent timing
resolution of these anti-coincidence detectors will then allow to cut
cosmic ray backgrounds to near negligible levels.

\begin{table}[ht]
\centering
\begin{tabular}{|c|c|c|c|c|c|}
\hline
   & CODEX-b & FASER$^{r}$ & FASER$^{R}$ & FASER$^n$   &  MATHUSLA      \\
\hline
$L_{min} (m)$   & 25  & 390  & 390   & 145    & 141 \& 269      \\
$L_{max} (m)$  & 35  & 400  & 400   & 150    & 170 \& 323      \\
\hline
$\phi$   & 0.4  & $2\pi$   & $2\pi $ & $2\pi$   & $\pi/2$      \\
$\eta$   & [0.2,0.6] & [8.3,$+\infty$] & [6.68,$+\infty$]  
    & [8.92,$+\infty$] & [0.88,1.65] \\
\hline
$\mathcal{L} (fb^{-1})$ & 300 & 3000 & 3000 & 3000  & 3000      \\
\hline
$f_D$ & $0.12$ & $0.009$ & $0.009$ & $0.012$ & $0.06$ \\
\hline
$\epsilon_{\text{geometric}} $ &  $0.01$  &  $6.25\times 10^{-8}$       
         &   $1.56\times 10^{-6}$ & $1.78\times 10^{-8}$   & $0.028$       \\
\hline
\end{tabular}
\caption{Summary of the different detector parameters. The FASER
  proposal \cite{Feng:2017vli} discusses three different possible
  setups: FASER$^{r}$ is the far detector with small radius $r=0.2$ m,
  while FASER$^{R}$ is the far detector with large radius $R=1$ m;
  FASER$^n$ is the near detector configuration with radius $0.04$ m. Here,
  $L_{min}$/$L_{max}$ are the minimum/maximum distance of the detector
  from the IP.  For MATHUSLA the distances at the near end and the
  far end are given separately. $\phi$: azimuthal angle
  coverage. $\eta$: pseudo-rapidity range covered by the
  detector. $\mathcal{L}$: luminosities used in the original
  papers. $f_D$ is the estimated maximum fraction of events decaying
  inside the detector volume. $\epsilon_{\text{geometric}}$: a naive
  estimate of the geometric acceptance of each detector, measured in
  fraction of the full solid angle ($4\pi$). Note, however, that FASER
  covers the extremely forward direction. Thus, in case of FASER the
  geometric acceptance underestimates the fraction of the total
  cross-section covered by a considerable factor, in particular for
  events from D- and B-mesons.}\label{TBL:detectors}
\end{table}

In Table.~\ref{TBL:detectors}, we summarize the relevant parameters of
the different experiments. $\mathcal{L}$ is the value of the
luminosity assumed in the original proposals.  Note that the authors
of \cite{Gligorov:2017nwh} give an expected luminosity of $300$/fb for
LHCb for the end of Run-5 of the LHC.
\footnote{According to \cite{Albrecht:2017odf} LHCb will take around
  50/fb until 2029. The LHCb collaboration has recently
  \cite{Aaij:2244311} expressed interest to run in a phase-II until
  2035, taking up to $300/$fb of data.}  This is a factor of 10 lower
than used in the other proposals, which make use of the high
luminosity environment in ATLAS or CMS.  We also give $f_D$, the
maximum fraction of events decaying within the detector length,
estimated for the optimal decay length, given the corresponding
$L_{min}$ and $L_{max}$. $\epsilon_{\text{geometric}} $ is the
geometric factor, calculated relative to the full solid angle.  Even
FASER$^R$ corresponds to an instumented volume of only roughly 8
m$^3$, while CODEX-b corresponds to $10^3$ m$^3$ and the massive
MATHUSLA covers a volume of $8\cdot 10^5$ m$^3$.  Note, however, that
FASER sits at very large $\eta$. Thus, $\epsilon_{\text{geometric}}$
underestimates the sensitivity of FASER for events coming from D- and
B-meson decays, see the discussion in the next section.

We note that in \cite{Kling:2018wct}, the authors discuss a slightly
modified setup for FASER with respect to those quoted in table
(\ref{TBL:detectors}). In particular they use a distance of
$L_{max}$=480 m to the IP. We will comment on the resulting
differences with respect to our calculation in section \ref{subsect:strl}.

We now briefly describe our simulation. We use \texttt{SARAH}-4.12.2
\cite{Staub:2013tta} to generate UFO \cite{Degrande:2011ua} models for
sterile neutrinos in a Seesaw Type-I model and the RPV-MSSM. We
generate spectrum files with \texttt{SPheno}-4.0.3
\cite{Porod:2011nf,Porod:2003um}. We created the Seesaw Type-I model
using the \texttt{SARAH} environment, while the RPV-MSSM model already
exists in the model repository.

For the W, Z and Higgs boson channels, we import Seesaw Type-I or
RPV-MSSM UFO models into \texttt{MADGRAPH5\_aMC@NLO} v2.6.0
\cite{Alwall:2014hca} where these bosons are generated, put on-shell
and decay. We then read the LHE \cite{Alwall:2006yp} event records
with \texttt{MadAnalysis5} v1.5
\cite{Conte:2012fm,Conte:2014zja,Jakovac:2014lqa}, which we use to
apply the relevant geometric cuts on the pseudo-rapidity. With these
simulated events we calculated the number of LLLPs produced in the
detector window ($n_f$) and the mean values of their
$\beta\gamma$. ($\langle \beta\cdot\gamma\rangle$ is needed for the
correct simulation of the decay length.) Dividing $n_f$ by the total
number of simulated events provides $f_{\text{window}}$, the fraction
of events relative to the total cross section. We prefer to use this
procedure, since we expect that the values of the total cross section
simulated in \texttt{MadGraph5} should have a larger uncertainty than
the relative fractions, as obtained in our procedure.

For the meson channels, we use the \texttt{HardQCD:hardccbar}(
\texttt{HardQCD:hardbbbar}) matrix element calculator with
\texttt{Pythia 8.205} \cite{Sjostrand:2006za,Sjostrand:2014zea} for
generating events $gg,q\bar{q}\rightarrow c\bar{c}(b\bar{b})$,
showering and hadronization. For simulating sterile neutrinos, we need
to know the fraction ($f_{\text{window}}$) of events of the total
cross section moving towards the different detectors and the mean
$\langle \beta\cdot\gamma\rangle$ of these windowed sterile
neutrinos. These can all be calculated and given by
\texttt{Pythia8}. In particular, \texttt{Pythia8} allows to add a
fourth neutrino to the standard model module, but does not
automatically recalculate branching ratios, when one varies the mass
of this fourth neutrino. For our sensitivity estimates we thus include
the phase space suppression of these branching ratios by hand, which
is important for non-zero sterile neutrino masses. We allow the B- and
D-mesons to three-body decay to the fourth neutrino plus an electron
and a lighter meson. Moreover, for a sterile neutrino mass close to
that of B- or D-mesons, the leptonic two-body decay of the charged
pseudoscalar mesons (in our case,
$\text{D}^{\pm},\text{D}_{\text{s}}^{\pm},\text{B}^{\pm},$ and
$\text{B}_{\text{c}}^{\pm}$) can lead to relatively large
contritbution. We therefore also include them in our calculation.

We then calculate the total number of events, using mostly
experimental data. For the D-meson and B-meson production we use the
results from LHCb
\cite{LHCb-CONF-2010-013,Aaij:2013mga,Aaij:2015bpa,Aaij:2016avz}.
LHCb gives cross sections within certain ranges of pseudo-rapidity
$\eta$ and $p_T$. We therefore use FONLL
\cite{Cacciari:1998it,Cacciari:2001td,Cacciari:2012ny,Cacciari:2015fta}
to extrapolate these cross sections to the full range of $\eta$ and
$p_T$. We checked that FONLL estimates roughly an uncertainty order 15
\% for the total cross section in case of D-mesons (and a similar
error for B-mesons).

Ref. \cite{Aad:2016naf} gives cross section of the $W^\pm \rightarrow
l^\pm \nu$ and $Z\rightarrow l^+ l^-$ production (where $l^\pm =
e^\pm, \mu^\pm$) in proton-proton collisions at $\sqrt{s}=13$
TeV. Making use of the branching ratios listed in the PDG
\cite{Patrignani:2016xqp}, we obtain the total cross sections for W
and Z bosons production (and also for the D- and B-mesons).  The Higgs
boson is produced at the LHC dominantly through gluon fusion. The
cross section can not be measured independently from the Higgs decay
branching ratios, thus the most reliable value for the total cross
section is probably still the calculated value of $43.92~pb$, as given
in \cite{deFlorian:2016spz}. With the total cross sections fixed this
way, we need only $f_{\text{window}}$ and $\langle
\beta\cdot\gamma\rangle$ from our simulation to calculate the total
number of events in the different detectors.

We calculate the decay width of the sterile neutrinos according to
formulas given in \cite{Atre:2009rg}. This should give a more accurate
treatment than the decay width calculated in \texttt{SPheno}, which
does not take into account hadronic form factors for decays to
mesons. We have checked that these form factors are numerically
important for sterile neutrino masses below 5 GeV. For the decay width
of the lightest neutralinos we take directly the output from the
\texttt{SPheno} spectrum files. \footnote{Since for the neutralino
  production and decay are not related, the exact calculation of the
  width is less important in this case.}

We calculate the number of LLLPs decaying inside the detector using the
following formula,
\begin{eqnarray}\label{eq:decN}
  N_{\text{decay}} = N_{\text{total}}~f_{\text{window}}~
  \big(e^{[-L_{\text{min}}/L_{\text{decay}}]}-e^{[-L_{\text{max}}/L_{\text{decay}}]}\big).
\end{eqnarray}
Here, $N_{\text{decay}}$ is number of the LLLPs decaying in the
detector, $N_{\text{total}}$ is the total number of LLLPs produced at
the IP, while $L_{\text{decay}}=\beta\gamma c \tau$ is their decay
length. $f_{\text{window}}$ is the fraction of events of the total
emitted into the detector window.  $L_{\text{min}}$ ($L_{\text{max}}$)
is the minimum (maximum) distance from the IP to the detector.  For
each detector, $N_{\text{total}}$ and $f_{\text{window}}$ are
functions of the mass of the LLLPs, and as described above, we
simulate these values using either \texttt{MadAnalysis5} or \texttt{Pythia8}.

The simple description given in eq. (\ref{eq:decN}) corresponds to
approximating the detectors as cones defined by their coverage in
$\eta$ and $\phi$. Since the orientation of MATHUSLA is not well
described by such a simple cone with a tip at the IP, we refine this
part of the calculation for MATHUSLA. We assume that
$f_{\text{window}}$ and $\langle\beta\cdot\gamma\rangle$ are constant
over the range of $\eta$ covered and then divide MATHUSLA into 10
smaller boxes of equal size. We then sum up the decay number of events
in each box to obtain the total number. We have tested this rather
simple approximation against a calculation done with a Mathematica
notebook, distributed by the authors of \cite{Chou:2016lxi}, which
integrates numerically over the detector valume. We find good
agreement between our simple-minded approach and the more accurate one
given in \cite{Chou:2016lxi}.

\section{Numerical results}
\label{sect:res}

\subsection{Sensitivities for different production modes}
\label{subsect:ctau}

Sterile neutrinos are produced via their mixing with the active
neutrinos. Thus, for sterile neutrinos production and decay width are
both governed by the same parameter, ie.  $|V_{\alpha N_j}|^2$.  The
situation is different for other LLLP candidates.  For example, as
discussed in section \ref{subsect:rpv}, the neutralino in RPV SUSY
can be singly produced via its mixing with the neutrino or produced in
pairs, via Z- or Higgs diagrams. Since one expects that RPV parameters
are small (since neutrino masses are small) also neutralino-neutrino
mixing is expected to be small. Thus, the main production mode for a
light neutralino should be Z-boson (and/or Higgs) boson decays.

For this reason, we will first give sensitivity estimates for the
different experiments separated into the different possible production
modes. We consider D-mesons, B-mesons, the gauge bosons W and Z, as
well as the Higgs.  Also, in this subsection we will present our
estimates in the particular parameter plane branching ratio versus
$c\tau$. This has the advantage that the results shown can be easily
used to estimate also sensitivities for other LLLP candidates.

\begin{figure}[ht]
\centering
\includegraphics[scale=0.42]{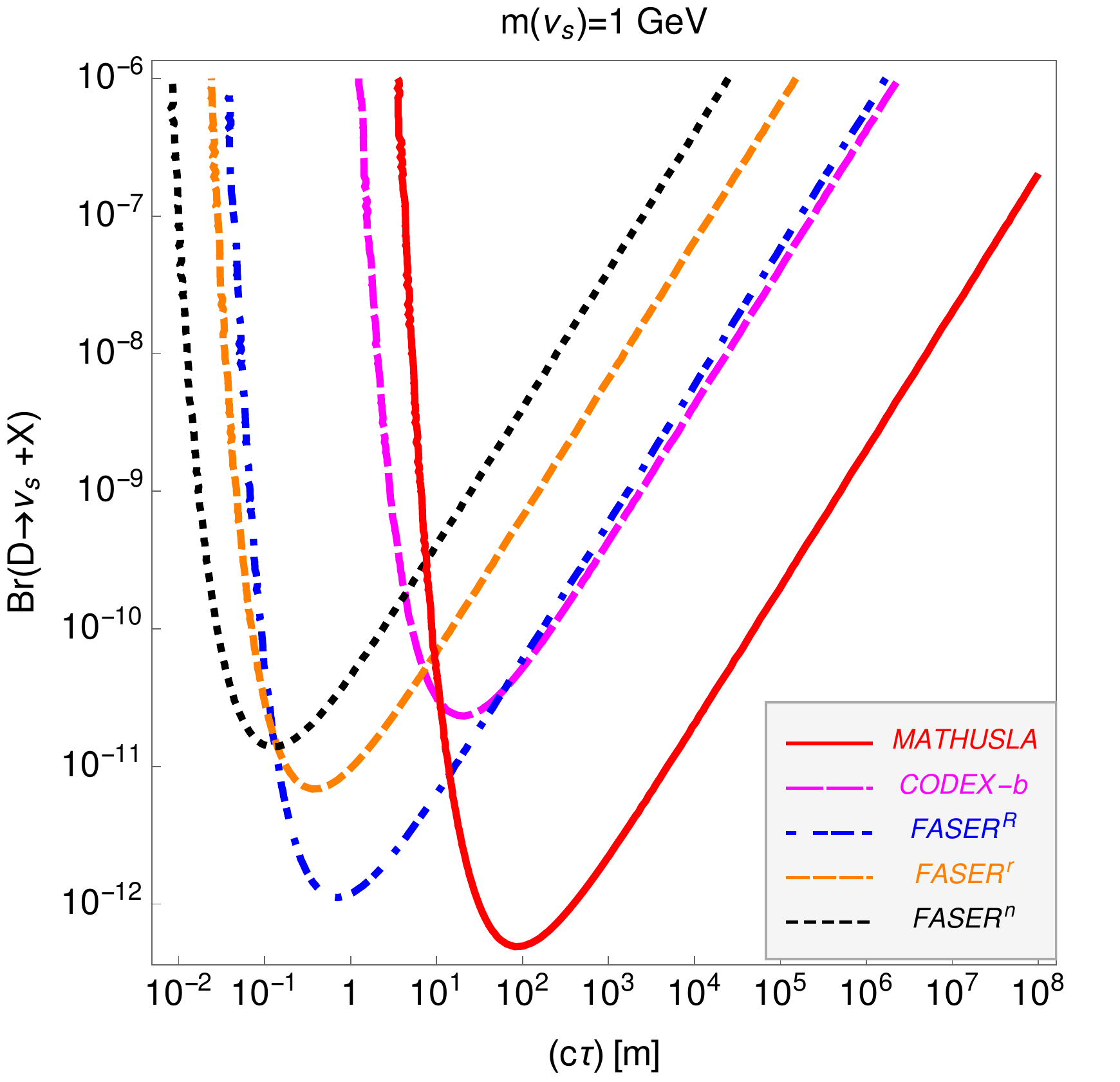}
\includegraphics[scale=0.42]{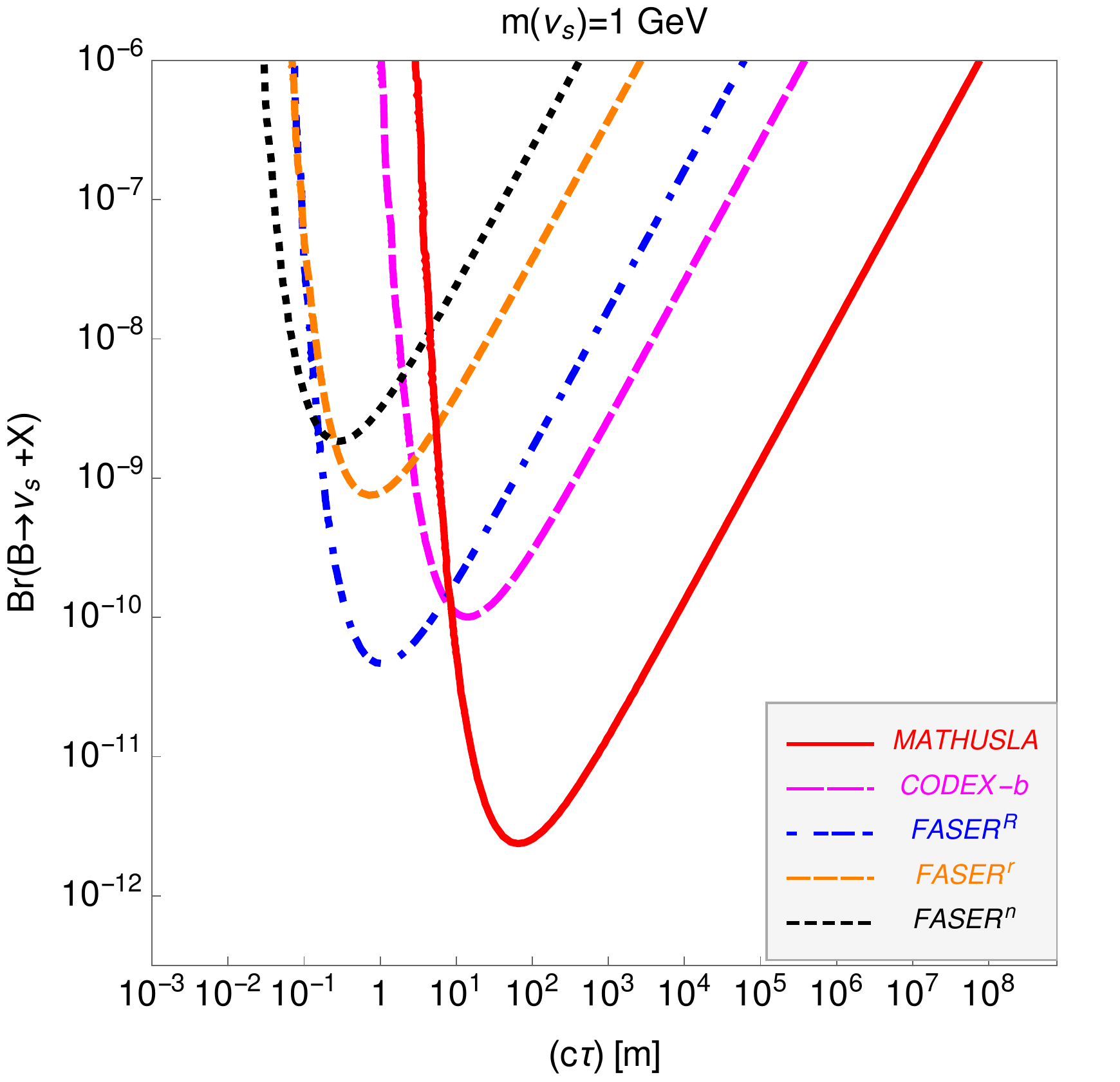}
\caption{Sensitivity estimates in the plane branching ratio versus
  decay length ($c\tau$) for CODEX-b, FASER and MATHUSLA.  The plot to
  the left is for neutral fermions from D-meson decays, to the right
  the corresponding results for B-mesons. Here and in all other
  sensitivity plots we use the contours of 4 signal events to estimate
  the sensitivity limit.}
\label{FIG:ctauBandD}
\end{figure}

In fig.~(\ref{FIG:ctauBandD}) we show the estimated reach of CODEX-b,
FASER and MATHUSLA in the plane branching ratio versus $c\tau$ for
D-mesons (left) and B-mesons (right). Here, and in all other plots in
this paper we use the contours of 4 signal events for estimating the
sensitivity limit.\footnote{This implies, of course, that we assume
  implicitly that the number of background events over the life-time
  of the experiments are smaller than this number.} As the plot shows,
FASER$^R$ and MATHUSLA are sensitive to different regions in $c\tau$,
despite being at similar distances from the IP.  This can be
understood, because mesons flying in the foward direction receive
typically much larger boosts than mesons produced more centrally. The
same effect makes FASER less sensitive at large $c\tau$. One notes
that at $c\tau \simeq 1$ m FASER$^R$ can reach nearly as small
branching ratios as MATHUSLA at $c\tau \simeq 10^2$ m in the case of
D-mesons, despite being a much smaller experiment. The reason is
simply that charm (and to a lesser degree bottom) quark production at
the LHC is very strongly peaked in the forward direction.  For the
same reason the different setups of the FASER experiment (for
parameters see table (\ref{TBL:detectors})) do particularly well for
D-mesons: while FASER$^r$ contains only 4 \% of the decay volume of
FASER$^R$, its sensitivity is only a factor of 6 smaller for events
coming from D-meson decays. FASER$^n$ is sensitive to smaller $c\tau$,
but is always less sensitive than the other FASER variants. CODEX-b is
less sensitive than either FASER$^R$ or MATHUSLA in case of meson
decays. However, this is partly due to the lower luminosity ($300/$fb
versus $3000$/fb) assumed for CODEX-b. Finally note that MATHUSLA does
significantly better than FASER$^R$ for B-mesons.

\begin{figure}[ht]
\centering
\includegraphics[scale=0.3]{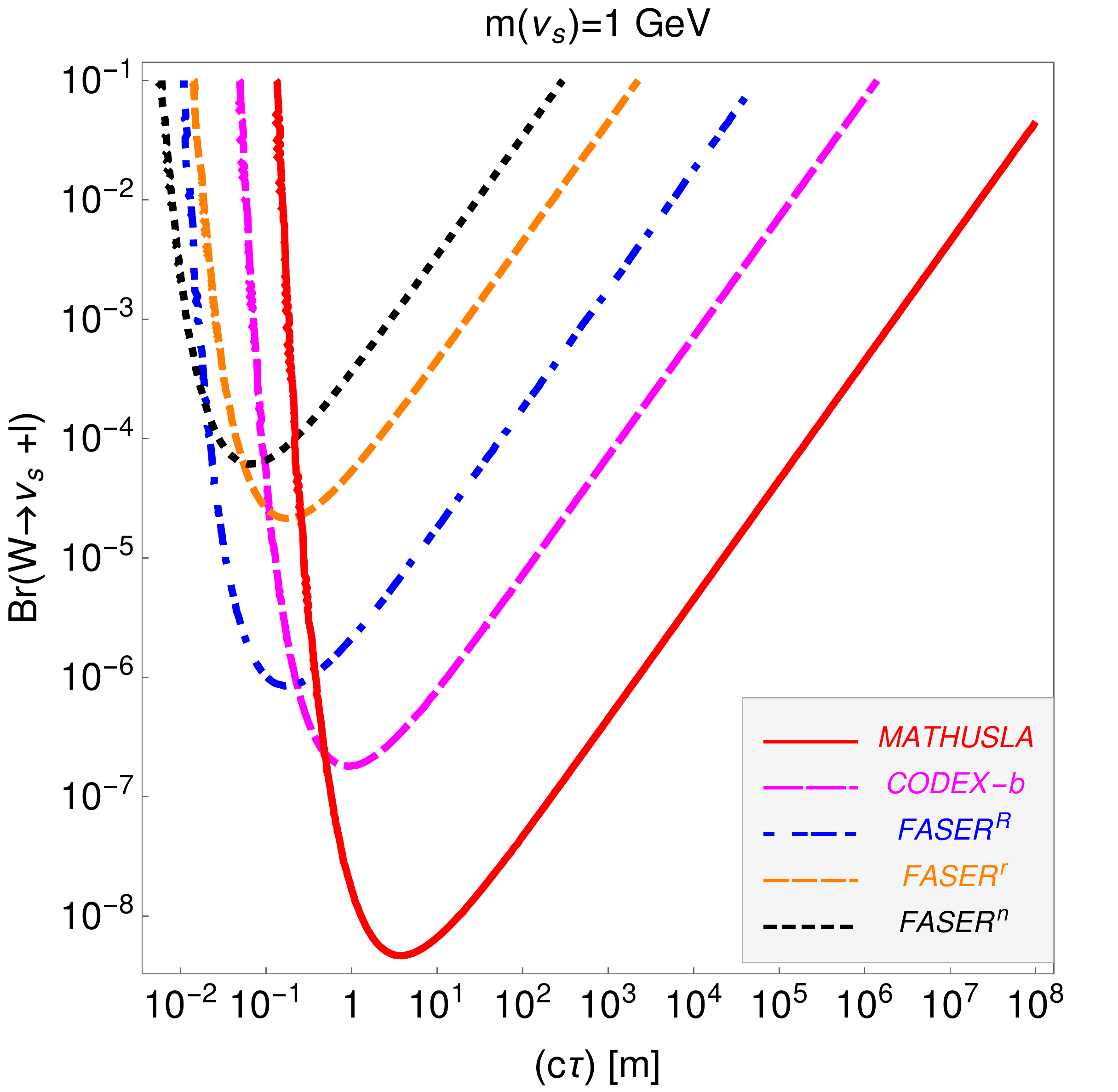}
\includegraphics[scale=0.3]{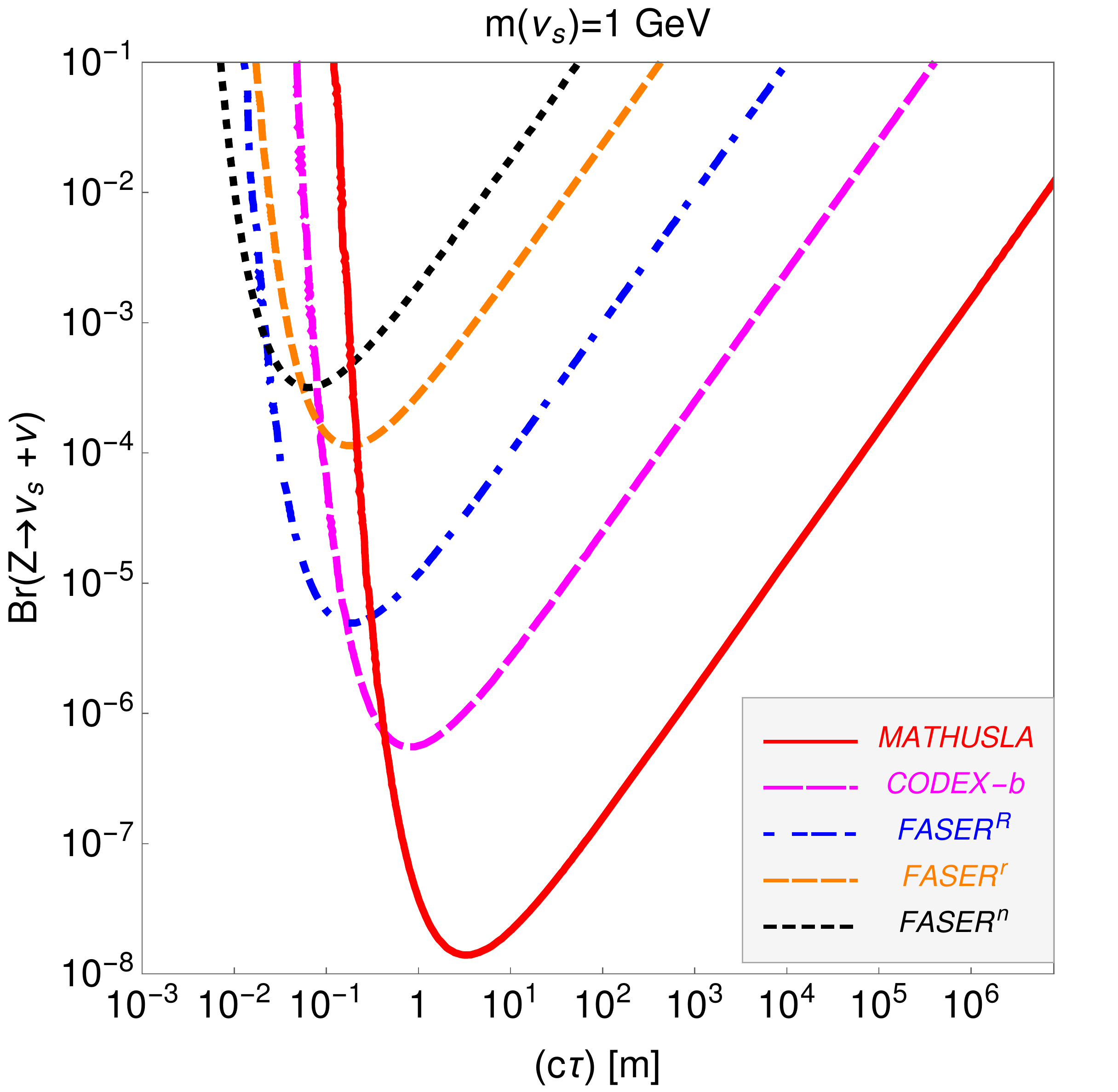}
\caption{Sensitivity estimates in the plane branching ratio 
versus decay length ($c\tau$) for CODEX-b, FASER and MATHUSLA. 
The plot to the left is for neutral fermions from W decays, 
to the right for Z-boson production.}
\label{FIG:ctauWandZ}
\end{figure}

In fig.~(\ref{FIG:ctauWandZ}) we show the results for production from
W- (left) and Z-bosons (right). Note the change of scale in the axis:
This simply reflects that there are 5 orders of magnitude more
D-mesons than gauge bosons produced at the LHC.  Comparing the results
shown in fig.~(\ref{FIG:ctauWandZ}) with those of
fig.~(\ref{FIG:ctauBandD}) one sees that CODEX-b now does better than
even FASER$^R$ for both, W and Z-bosons.  MATHUSLA is again the most
sensitive setup, apart from some region of parameter space at small
$c\tau$. Note also that FASER$^r$ and FASER$^n$ give much weaker
limits than FASER$^R$ in this case. The explanation is again simply 
that for gauge bosons there is much less boost into the forward
direction than for D-mesons.

\begin{figure}[ht]
\centering
\includegraphics[scale=0.3]{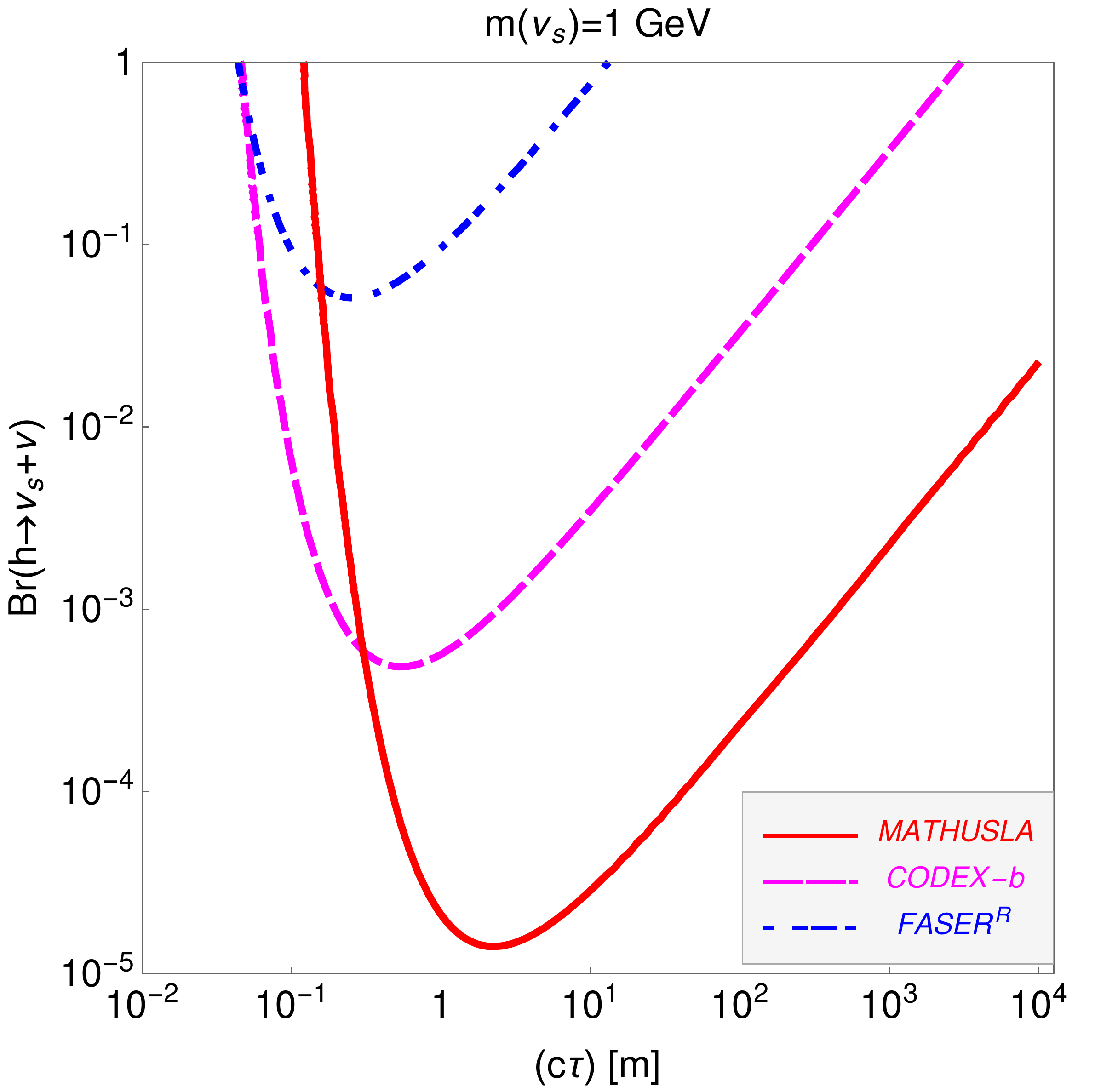}
\includegraphics[scale=0.3]{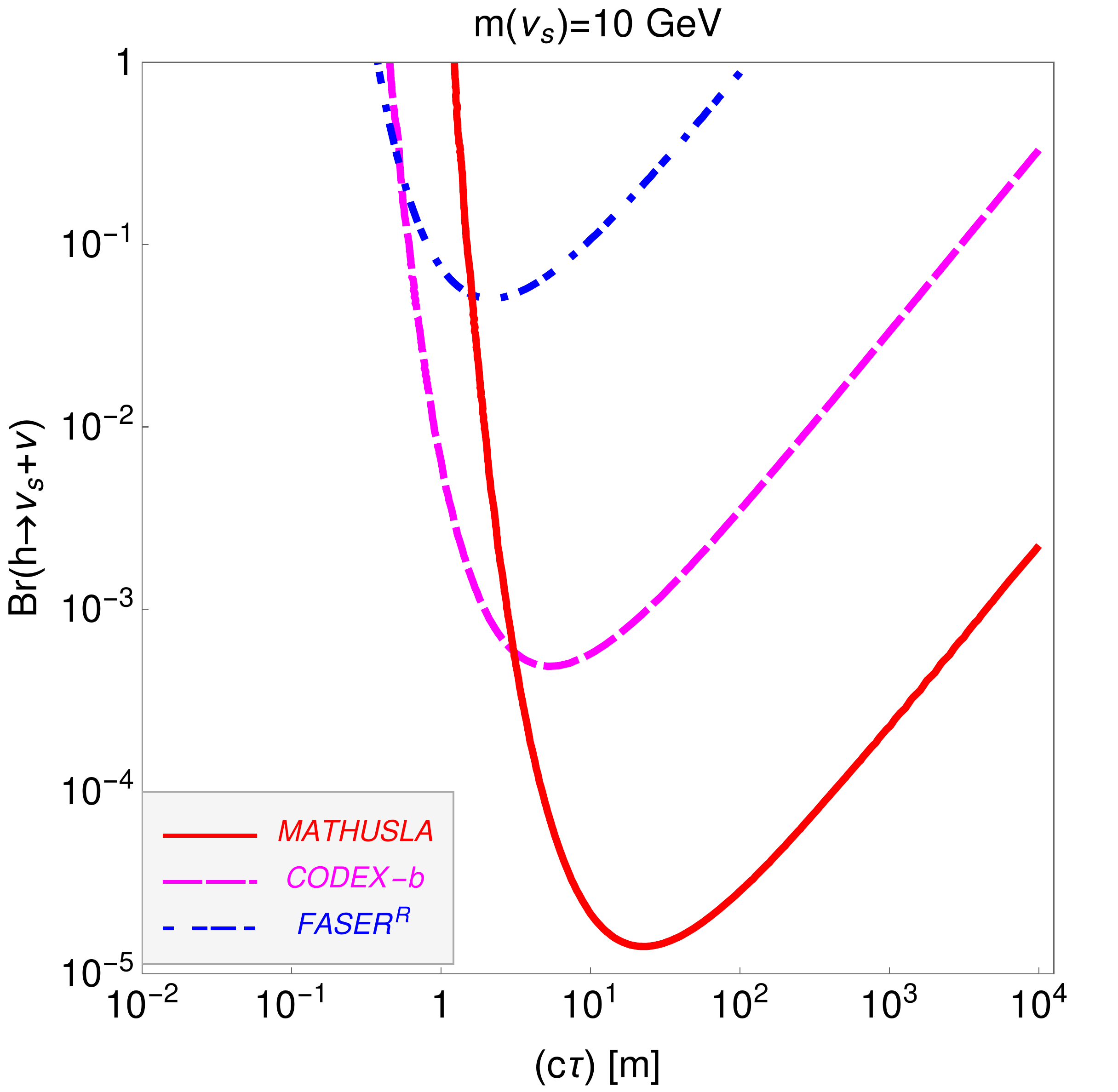}
\caption{Sensitivity estimates in the plane branching ratio 
versus decay length time ($c\tau$) for CODEX-b, FASER and MATHUSLA. 
The plot to the left is for neutral fermions with a mass of 1 GeV, 
to the right 10 GeV, both for Higgs production.}
\label{FIG:ctauHiggs}
\end{figure}

We have also calculated the corresponding expectation for the
different experimental setups for Higgs production, see
fig. (\ref{FIG:ctauHiggs}). The figure shows no contours
for FASER$^r$ and FASER$^n$ since these variants have no
sensitivity in the plane plotted here (even with $3000/$fb
of luminosity). Also FASER$^R$ can not compete with CODEX-b
or MATHUSLA in case of Higgs production. CODEX-b is around
a factor $\sim$ 40 less sensitive than MATHUSLA. However,
recall that (i) CODEX-b is a much smaller setup and (ii)
this estimate uses only $300/$fb for CODEX-b.

In fig.~(\ref{FIG:ctauHiggs}) we show the sensitivity ranges
for two choices of $m_{\nu_s}$. To the left we use  $m_{\nu_s}= 1$
GeV, while the plot on the right uses $m_{\nu_s}=10$ GeV. This
change of mass leads to a change in the value of $\langle\beta\cdot
\gamma\rangle$ and thus to a corresponding shift in the region
in $c\tau$, where the experiments are most sensitive. Since
this shift is similar for all the three experiments shown, however,
this does not affect the conclusions.
We note in passing that the curve for MATHUSLA on the right of
fig. (\ref{FIG:ctauHiggs}) is very similar to the corresponding one
shown in fig.~3 of \cite{Chou:2016lxi} for a scalar LLLP with the same
mass. 

\subsection{Sterile neutrinos}
\label{subsect:strl}

We now turn to a discussion of the sterile neutrino results.  For the
calculation of the sensitivities of the different experimental
proposals we take into account the different production processes
discussed in the previous subsection: D-meson and B-meson decays, W-
and Z-bosons, as well as Higgses. For the sterile neutrino case,
production of steriles from Higgs decays gives only a negligible
contribution to the sensitivity, as expected.

In our estimates we will not consider sterile neutrinos decaying to
$\tau$'s. We will also consider only mixing of the sterile neutrinos
with either $e$ or $\mu$ for simplicity. For this simplified case,
assuming the experimental detection efficiencies for $e$ and $\mu$ to
be similar, plots for $|V_{e N}|^2$ and $|V_{\mu N}|^2$ are very
similar and we simply will show plots using $|V_{\alpha N}|^2$.

\begin{figure}[ht]
\centering
\includegraphics[scale=0.42]{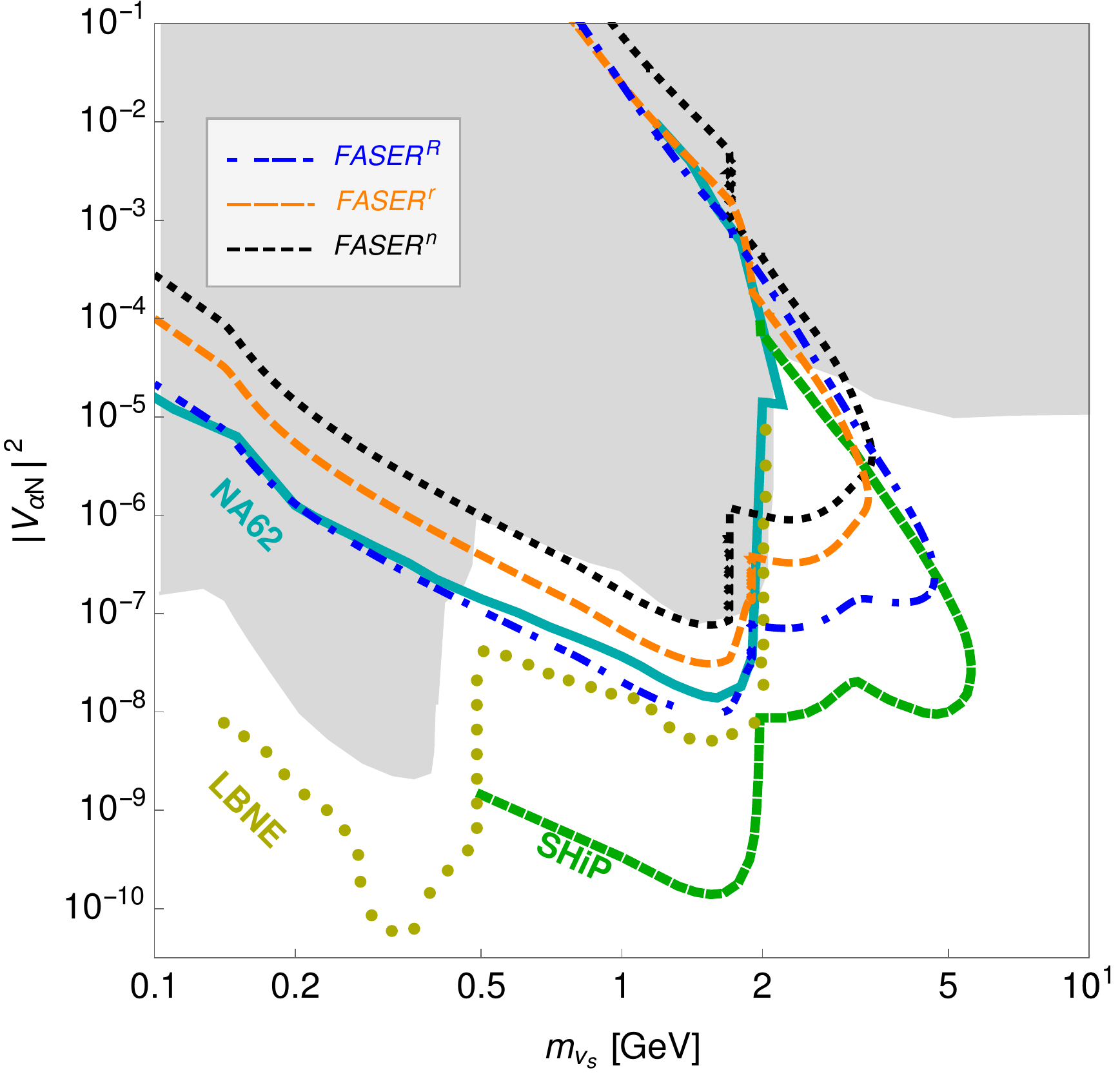}
\includegraphics[scale=0.42]{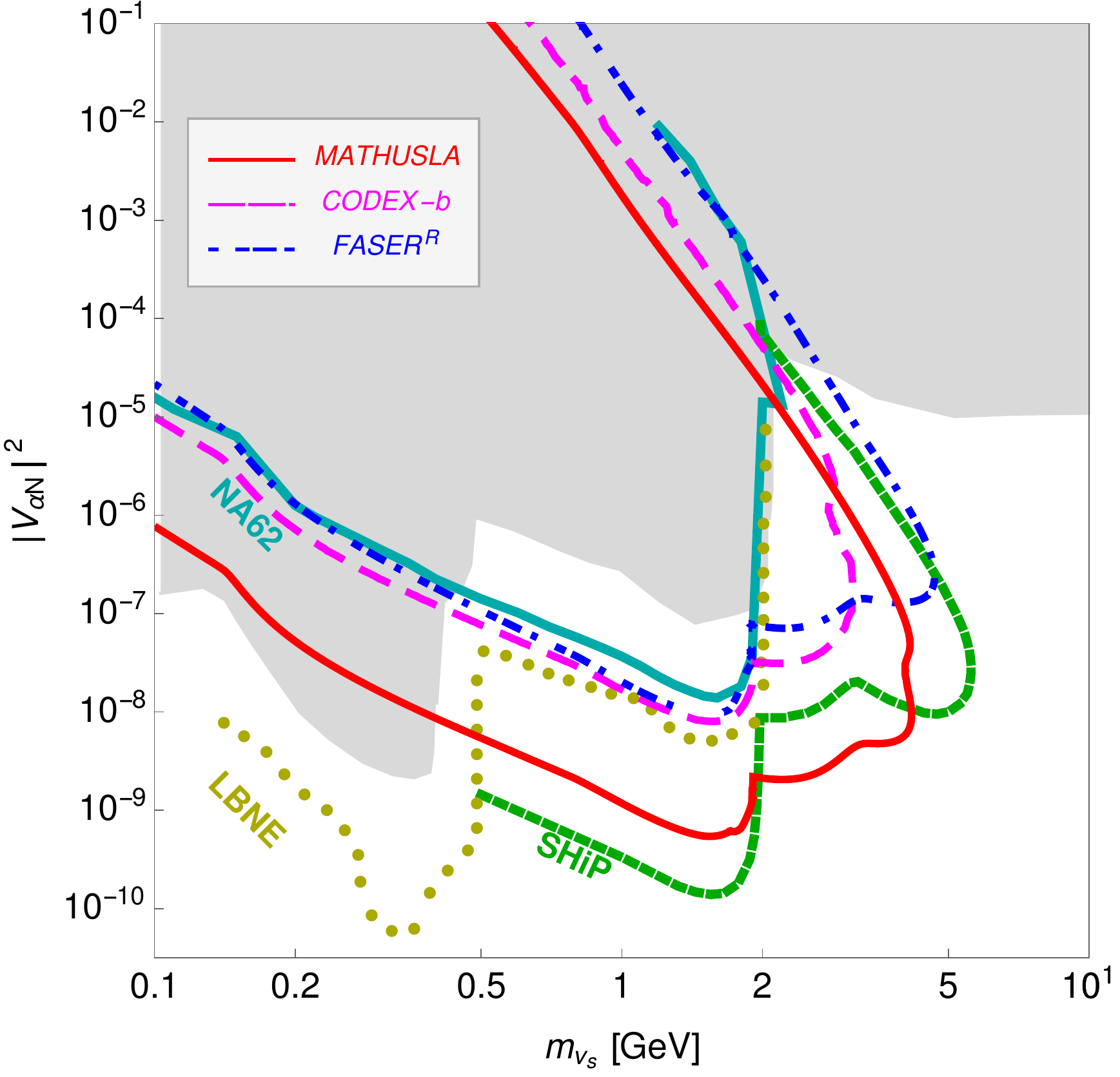}
\caption{Sensitivity estimates in the plane 
mixing angle squared, $|V_{\alpha N}|^2$ versus mass of the sterile 
neutrino, $m_{\nu_S}$ [GeV]. To the left we compare the different 
variants of FASER, to the right FASER$^R$, CODEX-b and MATHUSLA 
compared to LNBE and SHiP. For a discussion see text.}
\label{FIG:AllVsqMnuS}
\end{figure}

In fig. (\ref{FIG:AllVsqMnuS}) we show a comparison of the different
experiments in the plane $|V_{\alpha N}|^2$ versus mass of the sterile
neutrino, $m_{\nu_S}$ [GeV]. To the left we compare the different
variants of FASER to each other, while the plot on the right compares
FASER$^R$, CODEX-b and MATHUSLA. We also show the projected
sensitivity of SHiP \cite{Alekhin:2015byh,Boiarskyi:2018}, the
expectations for the near detector of the future DUNE experiment
\cite{Adams:2013qkq}, denoted as LBNE in the plot, and a recent
estimate for the final sensitivity of NA62 \cite{Drewes:2018gkc}. For
a description of the NA62 experiment see \cite{NA62:2017rwk}; for the
current status of limits for HNFs from NA62 see
\cite{Kholodenko:2017ecj}.  The grey area in the background is
excluded from past searches \cite{Deppisch:2015qwa}. The experimental
references used in drawing the excluded areas are PS191
\cite{Bernardi:1987ek}, JINR \cite{Baranov:1992vq}, CHARM
\cite{Bergsma:1985is} and DELPHI \cite{Abreu:1996pa}.

The plot on the left of fig. (\ref{FIG:AllVsqMnuS}) shows that, as
expected, FASER$^R$ always does better than the other FASER
variants. It also shows that below $m_{\nu_S} \simeq 2$ GeV FASER$^R$
is competitive with NA62 \cite{NA62:2017rwk}, while for $m_{\nu_S}
\gsim 2$ GeV, FASER has better sensitivity than NA62. For the whole
mass range, SHiP is more sensitive than FASER$^R$.

The plot on the right of fig. (\ref{FIG:AllVsqMnuS}) shows that
CODEX-b and FASER$^R$ have quite similar sensitivities below
$m_{\nu_S} \simeq 3.2$ GeV to sterile neutrino parameters, while
MATHUSLA does better than both in most parts of the parameter
space. While the plot shows that SHiP has the best expected
sensitivity in the mass range $0.5 \lsim m_{\nu_S} \lsim 2$ GeV, for
larger masses FASER$^R$ and CODEX-b are only worse than SHiP by
approximately one order of magnitude. One notes also that MATHUSLA is
only slightly less sensitive than SHiP for $m_{\nu_S} \lsim 2$ GeV,
and even better than SHiP for $2 \lsim m_{\nu_S} \lsim 4$ GeV. Below
$m_{\nu_S} \lsim 0.5$ GeV the most sensitive experiment will be LBNE
\cite{Adams:2013qkq}. Note, however, that our calculation does not
include sterile neutrinos from Kaon decays, which provide most of the
sensitivity of LBNE.

We also briefly compare the results of \cite{Kling:2018wct} with our
calculations. As mentioned in the introduction, \cite{Kling:2018wct}
estimates the sensitivity of FASER for sterile neutrinos. While our
results are overall similar to those shown in \cite{Kling:2018wct},
there are also some minor differences.  We believe this is mainly due
to the following: (i) \cite{Kling:2018wct} gives a decay length around
a factor of 2 larger than our calculation in this mass range; (ii) the
distance of FASER in \cite{Kling:2018wct} is set to 480 m (we use 400
m from the original proposal).

\subsection{The neutralino case}
\label{subsect:ntrl}

We now turn to a discussion of light neutralinos in R-parity violating
SUSY as LLLP candidates. In SUSY models with BRPV, neutralinos and
neutrinos mix. As discussed in section \ref{subsect:rpv}, this mixing
is related to the neutrino mass generated in these models. Constraints
on purely BRPV models for singly produced neutralinos can thus be
derived from a re-interpretation of the sterile neutrino constraints
discussed above.

The main phenomenological difference between sterile neutrinos and
neutralinos in RPV then actually comes from pair production of
neutralinos.  As discussed in section \ref{subsect:rpv}, pair
production of the lightest neutralino is not suppressed by small RPV
parameters.  We will consider neutralinos that are pair-produced from
Z-boson decays. At the LHC with $3000/$fb of luminosity there will be
a total of more than $10^{11}$ Z-bosons produced. The current upper
limit of Br($Z\rightarrow\chi^0_i\chi^0_j$) $\lsim 0.1$ \% at 90 \%
c.l. implies that at the maximum allowed still up to $10^8$
neutralinos could be produced from Z-decays.

\begin{figure}[ht]
\centering
\includegraphics[scale=0.3]{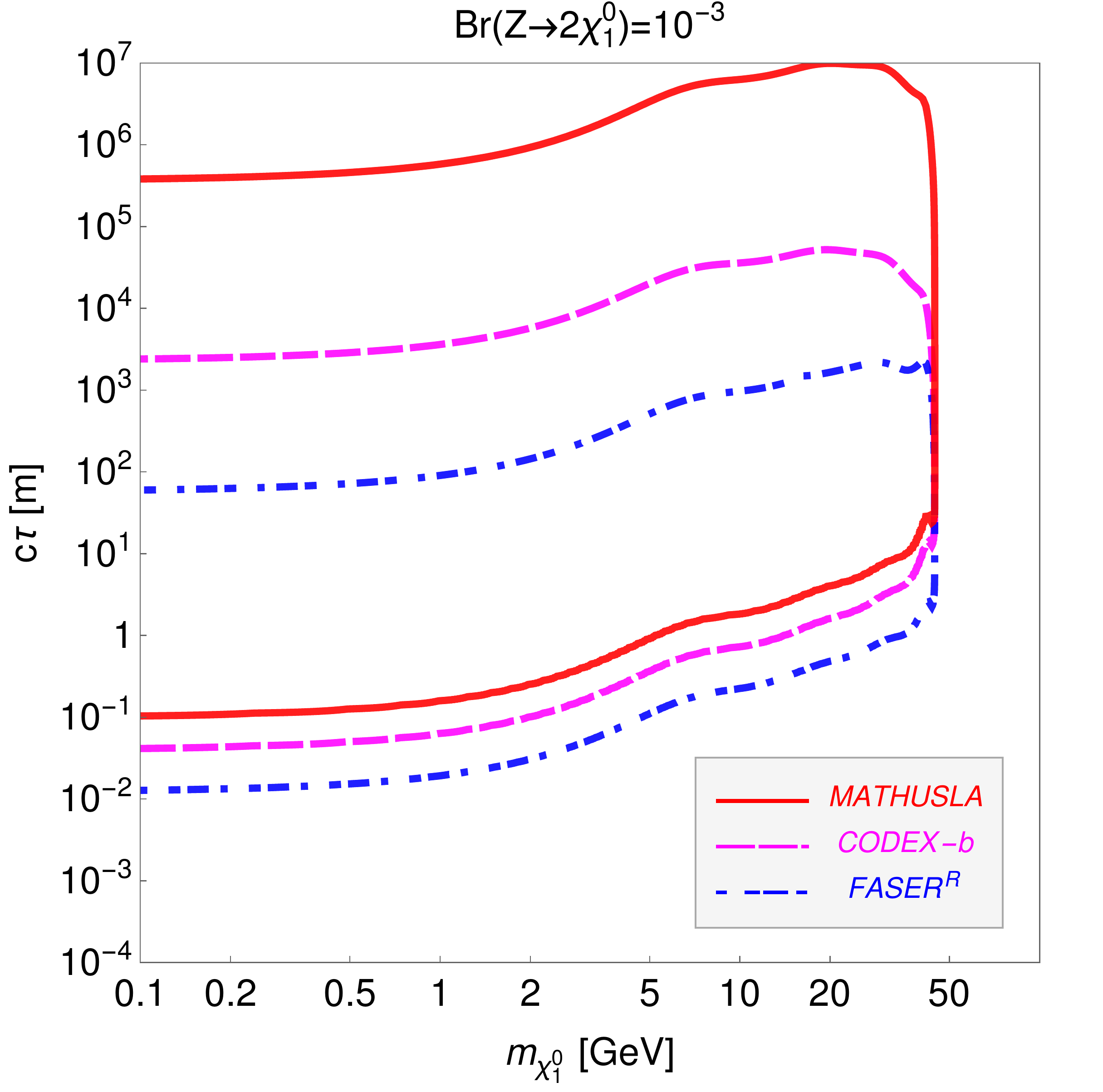}
\includegraphics[scale=0.3]{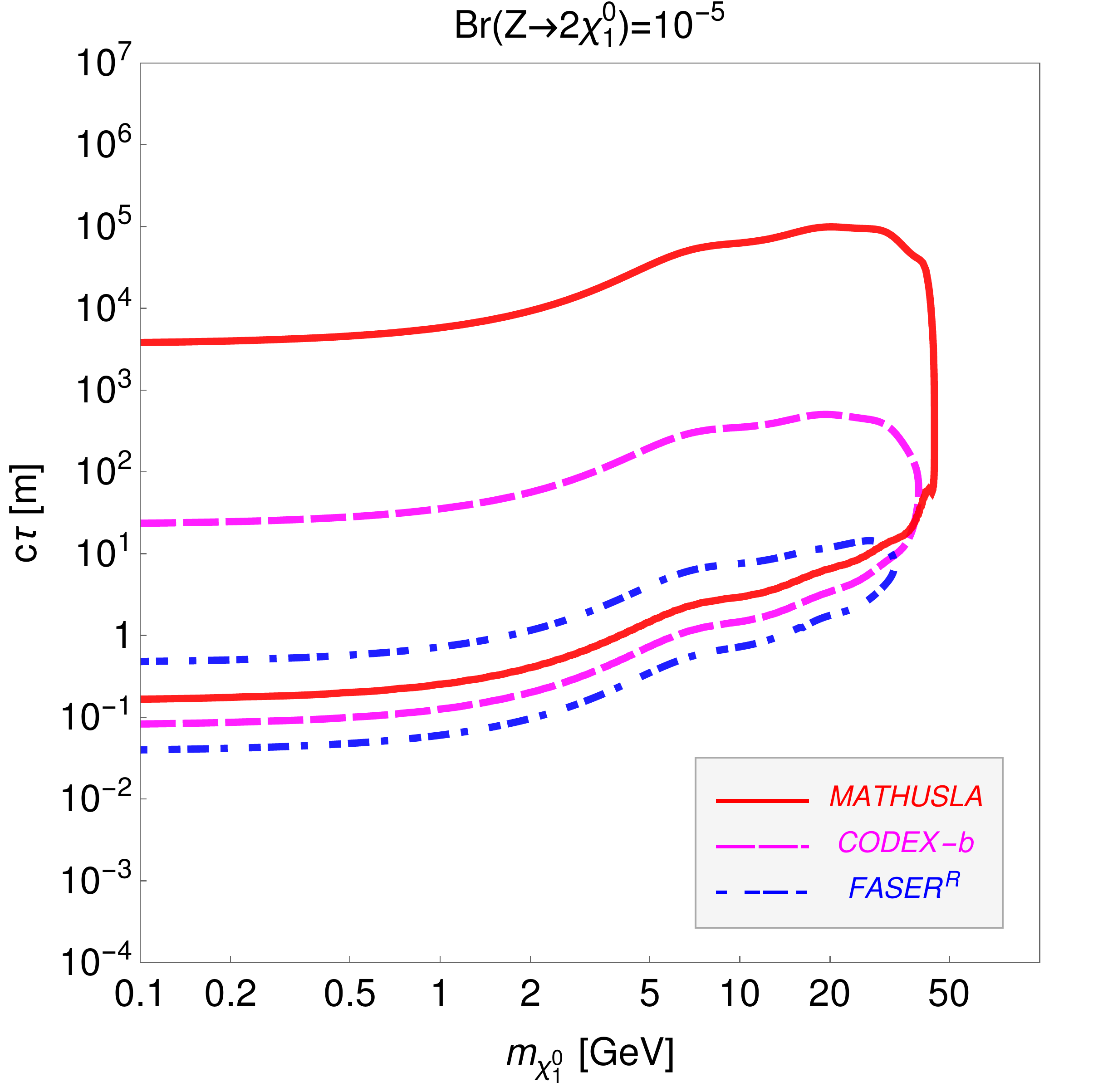}
\caption{Parameter space accessible for CODEX-b, FASER$^R$ and MATHUSLA
  for the lightest neutralino in R-parity violating SUSY. This
  calculation assumes neutralinos are pair-produced from Z-boson
  decays. The plot on the left assumes
  Br($Z\rightarrow\chi^0_1\chi^0_1$) $= 10^{-3}$ (i.e. a value
  saturating the experimental upper limit), the one on the right uses
  $10^{-5}$. }
\label{FIG:RPVctau}
\end{figure}

Fig. (\ref{FIG:RPVctau}) shows the accessible parameter space for
discovering events from neutralino decays in the plane $c\tau$ versus
lightest neutralino mass, $m_{\chi_1^0}$ [GeV].  Since we have only
upper limits on Br($Z\rightarrow\chi^0_1\chi^0_1$), we show two cases
in this figure. The plot on the left puts the branching ratio at the
experimental upper limit, while the plot on the right uses
Br($Z\rightarrow\chi^0_1\chi^0_1$) $= 10^{-5}$.\footnote{Here and
  everywhere else we refer to Br($Z\rightarrow\chi^0_1\chi^0_1$) as a
  number. This is to be understood as
  Br($Z\rightarrow\chi^0_1\chi^0_1$) for low values of $m_{\chi^0_1}$,
  i.e. $m_{\chi^0_1}\ll m_Z/2$. For $m_{\chi^0_1}$ approaching $m_Z/2$
  the decay is phase space suppressed and this suppression is taken
  into account in the calculation.}  As the figure shows, for
Br($Z\rightarrow\chi^0_1\chi^0_1$) $=10^{-5}$ the sensitivity of
FASER$^R$ gets significantly diminished. Roughtly for
Br($Z\rightarrow\chi^0_1\chi^0_1$) $= 5 \cdot 10^{-6}$ there will be
less than 4 events in FASER$^R$ for any combination of $c\tau$ and
$m_{\chi_1^0}$ from pair produced neutralinos. For FASER$^r$ and
FASER$^n$ the corresponding limits are $3\cdot 10^{-4}$ and $10^{-4}$,
respectively. Thus, the smaller variants of FASER could have
sensitivity only, if Br($Z\rightarrow\chi^0_1\chi^0_1$) is very close
to the current upper bound. For this reason, we do not show contours
for these two variants in fig. (\ref{FIG:RPVctau}).  For CODEX-b and
MATHUSLA the numbers are much more optimistic; we estimate these two
experiments can probe values down to
Br($Z\rightarrow\chi^0_1\chi^0_1$) $= 6 \cdot 10^{-7}$ and $1.4 \cdot
10^{-8}$ respectively.

As fig. (\ref{FIG:RPVctau}) also shows the parameter space testable in
FASER$^R$, CODEX-b and MATHUSLA covers the range of decay lengths
expected for such light neutralinos in SUSY models with bilinear RPV,
see also the discussion below eq. (\ref{eq:GamNtrlBRPV}) in section
\ref{subsect:rpv}. It is important to note that for pair produced 
neutralinos, a much larger mass range can be covered than in the 
sterile neutrino case. As the figure shows, neutralinos up to 
$m_Z/2$ can be tested. Again, this is due to the fact that production 
and decay of the neutralino are not related in pair production. 
Note also that, in particular, MATHUSLA can probe large
values of $c\tau$, which are interesting if (i) the neutralino is very
light and/or (ii) the SUSY parameters $M_2$ and/or $\mu$ are
large. Recall, however, that large values of $\mu$ automatically imply
a small higgsino content in $\chi^0_1$, leading to small values for
Br($Z\rightarrow\chi^0_1\chi^0_1$).

\begin{figure}[ht]
\centering
\includegraphics[scale=0.3]{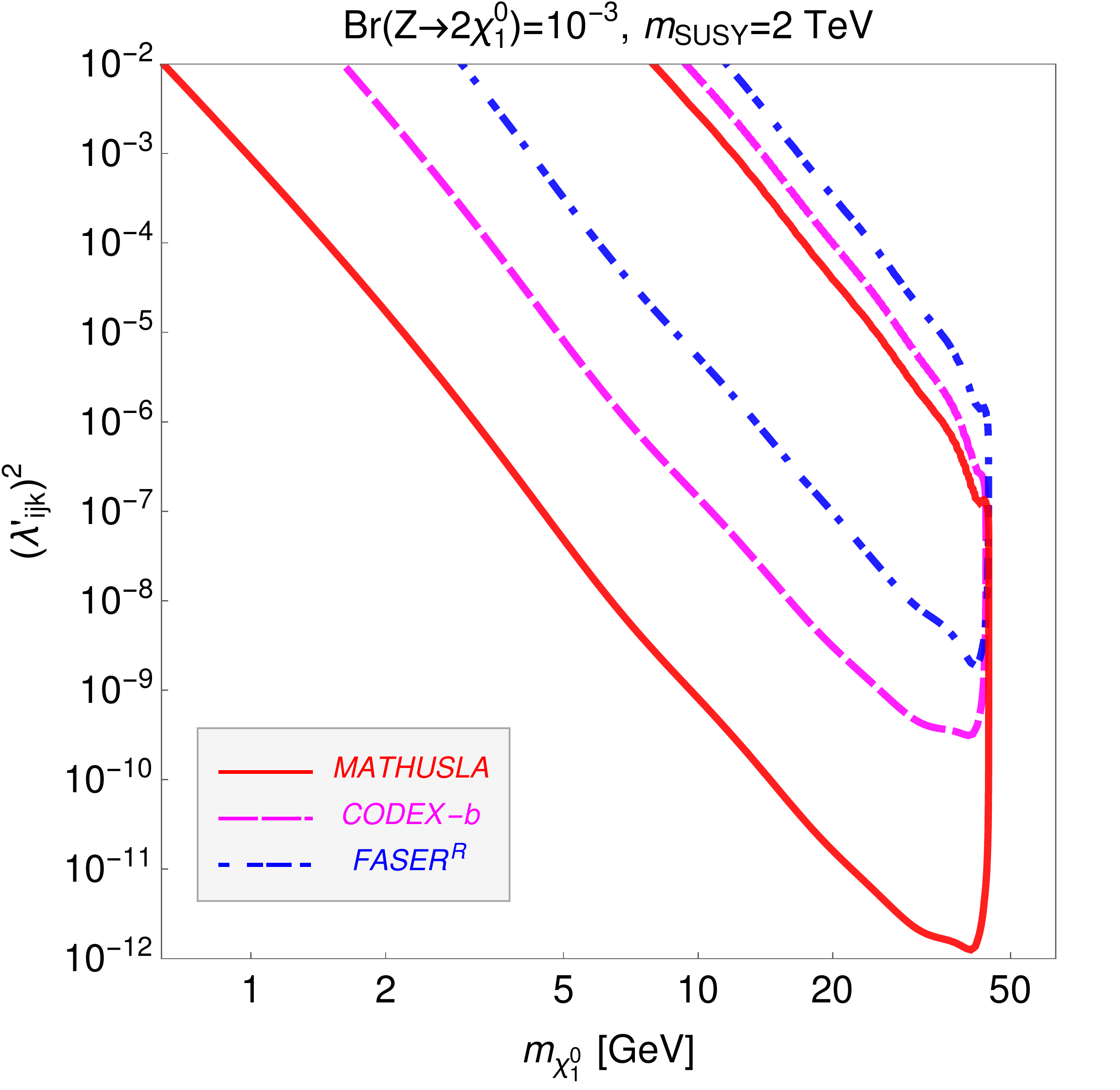}
\includegraphics[scale=0.3]{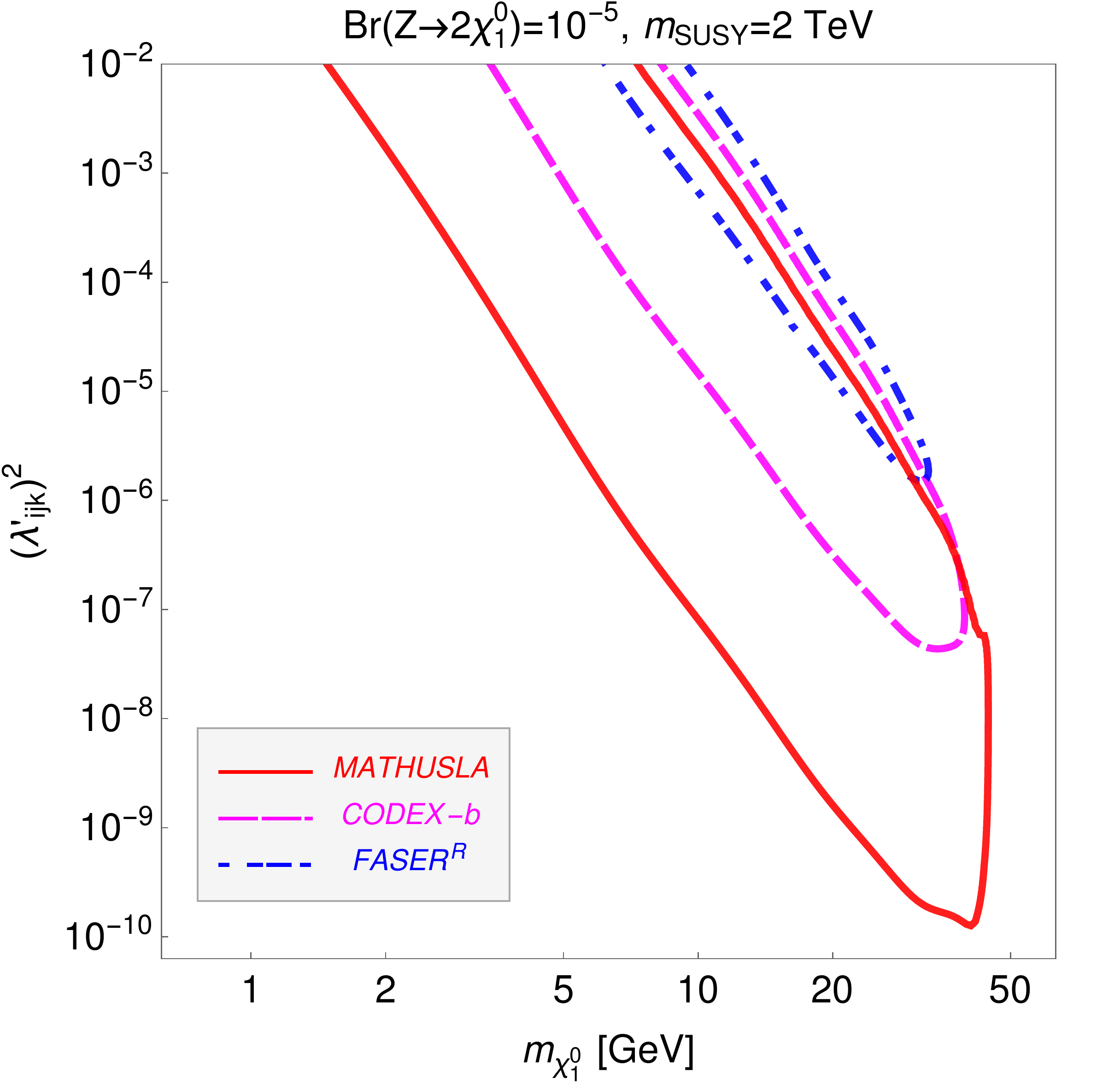}
\includegraphics[scale=0.3]{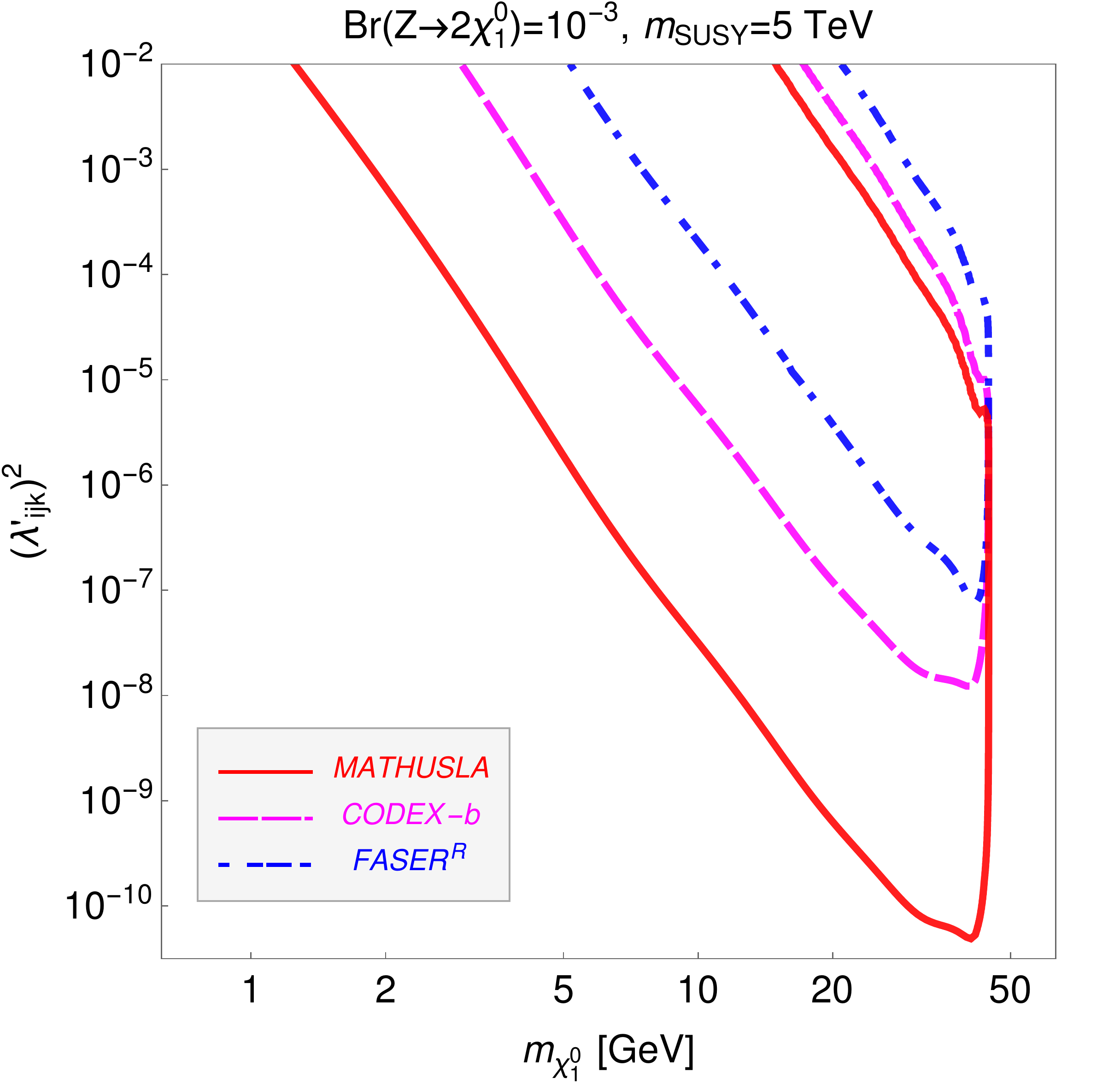}
\includegraphics[scale=0.3]{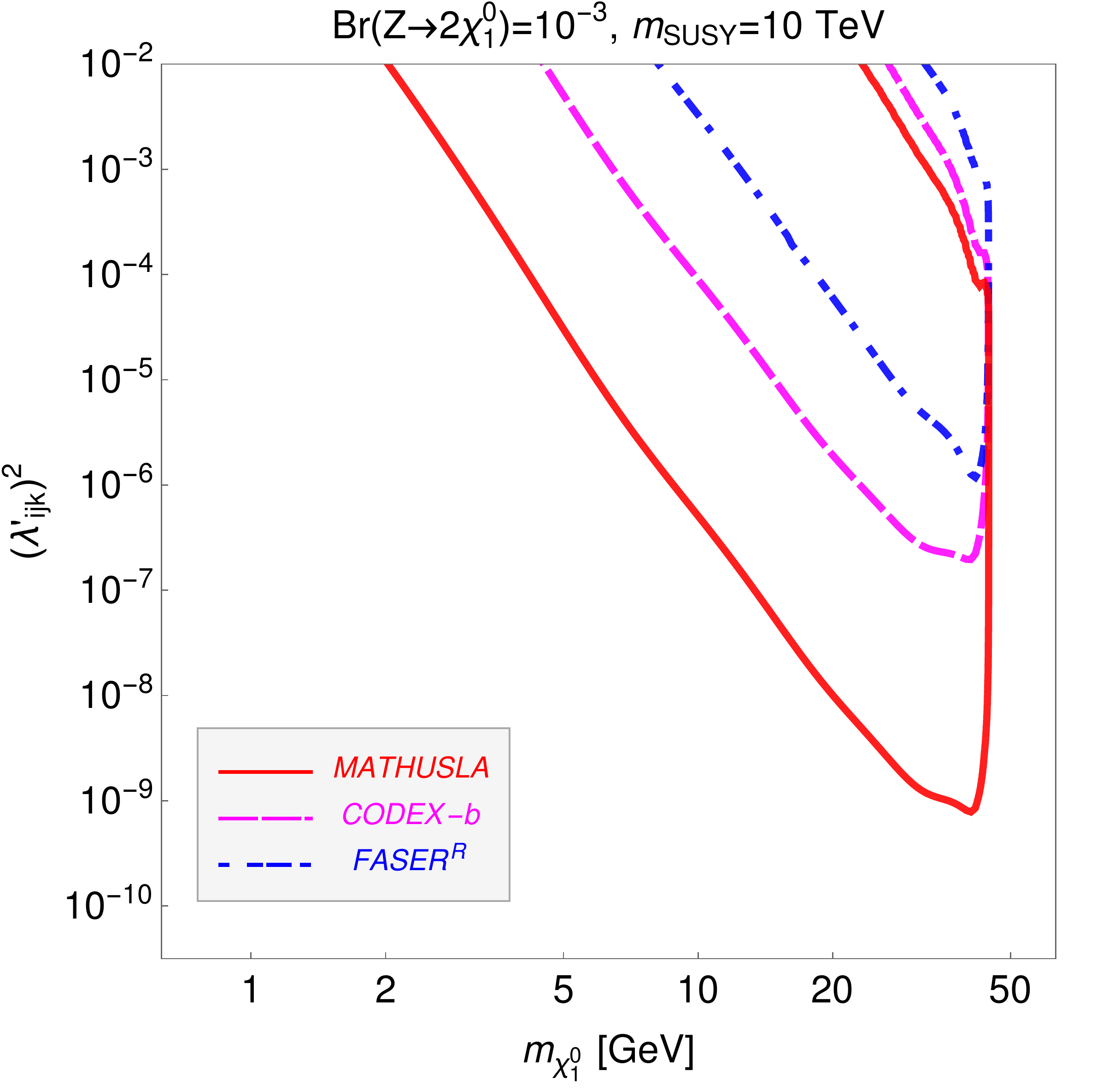}
\caption{Sensitivity estimates for CODEX-b, FASER$^R$ and MATHUSLA in
  the plane $\lambda'_{ijk}$ versus lightest neutralino mass.  In the
  top row we compare the sensitivities for two different values of
  Br($Z\rightarrow\chi^0_1\chi^0_1$) . The bottom row shows the
  change in reach for larger values of scalar masses, see text.}
\label{FIG:RPVlam}
\end{figure}

We now turn to a discussion of the case of trilinear RPV.  In
fig. (\ref{FIG:RPVlam}) we compare the different experiments in the
plane $\lambda'_{ijk}$ versus lightest neutralino mass for different
values of other parameters. Here, for $\lambda'_{ijk}$ the generation
indices $i,j,k$ could take in principle any value $1,2,3$, depending
on the final state discovered. However, in practice the results shown
are valid only for the first two generations of quarks and leptons,
since we do not take into account phase space suppression due to
non-zero final state quark and lepton masses in our calculation.  The
top row shows the sensitivities for two different values of
Br($Z\rightarrow\chi^0_1\chi^0_1$). Here, we have fixed all scalar
masses (both sleptons and squarks) to $m_{SUSY} \simeq 2$ TeV. The two
plots in the bottom row show how the explorable regions change for
larger values of the sfermion masses.  In these plots we assume that
the decay length is dominated by trilinear RPV and neglect any
possible contribution from BRPV.

Since the smallness of $\lambda'_{ijk}$ controls only the decay
length, but not the production cross section, in principle very small
values of $\lambda'_{ijk}$ are accessible in these searches. Note that
the values of $\lambda'_{ijk}$ shown can reach values several order of
magnitudes smaller than even the best of the current current upper
limits on $\lambda'_{ijk}$ \cite{Barbier:2004ez,Dreiner:1997uz}.
However, recall that pair production depends on the unknown value of
Br($Z\rightarrow\chi^0_1\chi^0_1$) , as discussed above.
Again, FASER$^R$ is expected to be less sensitive than CODEX-b,
with the best limits expected from MATHUSLA. 

In summary, if the lightest neutralino has a mass in the range (few)
GeV to $m_{Z}/2$, LLLP searches at FASER, CODEX-b and MATHUSLA can
probe part of RPV SUSY parameter space not accesible in any other
experiment. In particular, for bilinear RPV MATHUSLA can cover large
part of the predicted range of decay lengths for such light
neutralinos.
\section{Conclusions}
\label{sect:cncl}

We have discussed the sensitivities for three recent experimental
proposals, MATHUSLA \cite{Chou:2016lxi}, CODEX-b
\cite{Gligorov:2017nwh} and FASER \cite{Feng:2017uoz} for the case of
fermionic light long-lived particles. We considered two concrete
example models for our study: (a) light sterile neutrinos and (b) the
lightest neutralino in R-parity violating supersymmetry. Both
candidates are motivated by theoretical models that can explain the
observed small neutrino masses.

For sterile neutrinos, FASER$^R$ and CODEX-b show similar
sensitivities. Here, FASER$^R$ compensates its smaller detection
volume by taking advantage of the fact that D-mesons (and to some
degree B-mesons) are produced mostly in the forward
direction. MATHUSLA is more sensitive than either FASER$^R$ or CODEX-b
and, in fact, is competitive with the fixed target experiment SHiP.

For the case of neutralinos in RPV SUSY we have found that FASER$^R$,
CODEX-b and MATHUSLA can cover interesting parts of the parameter
space of these models, if the lightest neutralino has a mass in the
range of a few GeV up to $m_Z/2$. In particular, for bilinear RPV
models these experiments cover large parts of the range of $c\tau$
predicted theoretically from the observed neutrino masses. For
trilinear RPV, on the other hand, we have shown that if such a light
neutralino exists, RPV couplings can be probed which are orders
of magnitude smaller than all existing constraints.

Quite generally, MATHUSLA \cite{Chou:2016lxi} shows better sensitivity
to fermionic LLLPs than either CODEX-b \cite{Gligorov:2017nwh} or FASER
\cite{Feng:2017uoz}. However, this advantage clearly comes at a price:
MATHUSLA has by far the largest instrumented volume of all three
experiments. Considering that the variants of FASER, discussed so far
in the literature \cite{Feng:2017uoz,Kling:2018wct} are actually
quite small, compared to the other experiments, we think it would
be very interesting to study, if space for a larger version of FASER
in the very forward direction could be found.

\acknowledgments 

This work was supported by the Spanish MICINN grants FPA2017-85216-P,
SEV-2014-0398 and PROMETEOII/2014/084 (Generalitat
Valenciana). J.C.H. is supported by Chile grants Fondecyt No. 1161463,
Conicyt PIA/ACT 1406 and Basal FB0821. Z.S.W. is supported by the
Sino-German DFG grant SFB CRC 110 "Symmetries and the Emergence of
Structure in QCD". We thank Felix Kling for useful discussions on the
FASER proposal, and thank Jong Soo Kim and Torbj\"{o}rn Sj\"{o}strand
for help with the simulation softwares. We also thank Alex Pearce and
Patrick Spradlin for discussions concerning the LHCb experiment.  We
thank Marco Drewes, Jared Evans, Brian Batell and Alexey Boiarskyi for
discussing their results with us prior to publication.  Z.S.W. thanks
the IFIC for hospitality, and thanks the COST Action CA15108 for the
financial support during his research stay at the IFIC.

\bigskip

\end{document}